\documentclass[fleqn, 10pt, a4paper]{article}
\usepackage{float, graphics, lineno, color, authblk, multirow, makecell, fancyhdr, lastpage, booktabs}
\usepackage{newtxtext,newtxmath,amsmath}
\usepackage{orcidlink}
\usepackage[round]{natbib}
\usepackage[left=1.5cm, right=1.5cm, top=2.0cm, bottom=2.0cm]{geometry}
\usepackage[labelfont={bf,sf}, labelsep=period, justification=raggedright]{caption}

\hypersetup{colorlinks=true, allcolors=blue}
\DeclareCaptionLabelSeparator{sep}{$|$}
\makeatletter
\renewcommand\normalsize{%
\@setfontsize\normalsize\@xpt\@xiipt
\abovedisplayskip 2.5\p@ \@plus2\p@ \@minus5\p@
\abovedisplayshortskip \z@ \@plus3\p@
\belowdisplayshortskip 6\p@ \@plus3\p@ \@minus3\p@
\belowdisplayskip \abovedisplayskip
\let\@listi\@listI}
\makeatother

\title{\textbf{Do Candlestick Charts Suffice? The Incremental Economic Value of Technical Indicators in Stock Return Prediction}}
\author[1,*]{Peng Liu (pengliu@xaufe.edu.cn) \orcidlink{0000-0002-9115-2055}}
\author[2]{Yanyan Zheng (yanzheng@whu.edu.cn) \orcidlink{0000-0002-1092-072X}}

\affil[1]{School of Information, Xi'an University of Finance and Economics, Xi'an 710100, Shaanxi, P.R. China}
\affil[2]{School of Management, Xi'an Polytechnic University, Xi'an 710048, Shaanxi, P.R. China}
\affil[*]{Corresponding author: \href{mailto:pengliu@xaufe.edu.cn}{pengliu@xaufe.edu.cn}, \href{mailto:pengliuhep@outlook.com}{pengliuhep@outlook.com}}

\date{}

\begin{document}
\maketitle
\thispagestyle{empty}

\vspace{8mm}
\begin{abstract}
\noindent
Although technical analysis remains widely used in practice,
the need for manually constructed indicators is increasingly questioned as deep learning models can extract features directly from OHLC price data.
This paper examines the standalone economic value of trading volume and manually constructed technical-indicator groups and
whether adding these groups to OHLC candlestick charts provides incremental economic value.
Using daily data for 1,707 Chinese A-share stocks from 2013 to 2025,
we construct approximately 3.36 million stock-window samples and encode them as multi-channel images containing
candlestick charts, trading volume, and seven widely used manually constructed technical indicators.
The primary analysis uses a 2D CNN on these images;
complementary tests apply a logistic regression, a 1D CNN, and an LSTM to numerical representations of the same samples.
Out-of-sample performance from 2021 to 2025 is evaluated using decile portfolios, Sharpe ratios, Fama--French alphas, and formal pairwise tests.
Our empirical results yield two central findings.
First, specifications based separately on OHLC, trading volume, and each technical-indicator group produce significantly positive high-minus-low returns and factor-adjusted alphas,
together with economically meaningful Sharpe ratios;
the OHLC-only specifications generally perform best under the nonlinear models.
Second, adding these technical-indicator groups to OHLC does not systematically improve any of these performance measures in our setting.
Together, these results provide the first large-scale evidence from the Chinese A-share market on
the standalone and incremental economic value of these technical-indicator groups across multiple representations and model architectures.

\vspace{0.5em}

\end{abstract}

\noindent \rule{\textwidth}{0.8pt}

\section{Introduction}
\label{sec:intro}
Financial markets are widely viewed as complex adaptive systems in which prices emerge from interactions among heterogeneous,
boundedly rational investors whose beliefs and strategies evolve with market conditions \citep{Lo2004,SchollCalinescuFarmer2021,ZhaoLi2022}.
These interactions can generate nonlinear and time-varying dynamics,
while linkages among securities mean that individual stock returns arise within an interconnected market rather than from isolated processes \citep{Mantegna1999,Liu2025,Liu2026}.
Extracting stable predictive relationships from such an environment is therefore inherently difficult.

This complexity is also reflected in the statistical properties of asset returns.
Return distributions exhibit heavy tails and deviate substantially from Gaussian benchmarks,
implying that extreme price movements occur more frequently than standard distributional assumptions would suggest \citep{Mandelbrot1963,Mantegna1995,Cont2001,Liu2022}.
At short horizons, potential predictive signals must be extracted from an environment dominated by noise, dependence, and occasional large fluctuations.
Establishing economically meaningful return predictability therefore requires evidence that such signals persist out of sample and translate into portfolio performance.

Research on return predictability has long considered whether historical price and trading information contains signals about future returns.
Although the Efficient Market Hypothesis (EMH) maintains that prices reflect available information \citep{Fama1970},
complete and immediate information incorporation need not occur when information acquisition is costly, investors are boundedly rational,
or information diffuses gradually through the market \citep{GrossmanStiglitz1980,Lo2004,HongStein1999}.
More broadly, behavioral-finance research shows that investor beliefs, sentiment, and psychological states can
affect asset prices and perceptions of risk and return \citep{BarberisShleiferVishny1998,CanLeYu2024,LonkaniEtAl2025}.
Within this debate, technical analysis occupies a distinctive and controversial position.
Despite its widespread use among practitioners \citep{Menkhoff2010},
its academic standing remains unsettled.
Theoretical studies explain why historical prices may retain predictive content \citep{BrownJennings1989,HongStein1999},
and empirical studies document predictability based on past returns, technical signals, and price patterns across particular markets and asset classes
\citep{JegadeeshTitman1993,LoMamayskyWang2000,NeelyRapachTuZhou2014,HanZhouZhu2016,DetzelEtAl2020,JiangTongSong2019,JiangKellyXiu2023}.
Other studies, however, find that the economic performance of technical rules can be sensitive to rule selection, data snooping,
transaction costs, and risk adjustment \citep{SullivanTimmermannWhite1999,BajgrowiczScaillet2012,EugsterUhl2023}.
As \citet{LoMamayskyWang2000} observe, the enduring appeal of technical analysis may be partly attributable to its intuitive visual form.

Return predictability has also been studied using statistical and econometric models with prespecified predictors and relatively structured functional forms.
Linear predictive regressions assess whether lagged variables forecast subsequent returns and
whether the resulting forecasts improve on simple historical benchmarks out of sample \citep{WelchGoyal2008,CampbellThompson2008}.
Nonparametric methods relax some functional-form assumptions and have also been used to identify technical price patterns \citep{LoMamayskyWang2000}.
These approaches remain valuable for inference and interpretation,
but their performance can depend on the researcher's choice of predictors and functional form.
This dependence provides a natural motivation for more flexible data-driven methods.

Recent advances in machine learning and supervised statistical learning provide new tools for revisiting this longstanding debate.
In empirical asset pricing, tree-based ensemble methods, supervised dimension-reduction techniques,
and neural networks have been used to extract predictive signals from rich information sets,
with evidence of improved out-of-sample return forecasts relative to conventional benchmarks
\citep{GuKellyXiu2020,LeippoldWangZhou2022,XuPengSong2026}.

Within this broader development, deep learning has emerged as a prominent approach to stock market forecasting \citep{review2025-1,review2025-2}.
Studies using numerical market histories employ deep feedforward networks,
recurrent architectures such as long short-term memory (LSTM) networks,
one-dimensional convolutional neural networks (1D CNNs),
attention-based models, and hybrid architectures,
with prediction targets ranging from future price levels and directions to stock returns \citep{thakkar2021}.
Empirical applications across different markets often report strong predictive performance \citep{hiransha2018,FischerKrauss2018,gulmez2023,tran2024}.
However, these findings should be interpreted in light of differences in prediction targets and evaluation designs.
In particular,
high accuracy in forecasting short-horizon price levels may partly reflect persistence in the price series and
need not translate into economically valuable return forecasts \citep{Radfar2025}.

A complementary line of research represents market histories as visual charts, particularly candlestick charts.
Although candlestick images contain the same underlying opening, high, low, and closing (OHLC) observations as numerical series,
they organize price histories spatially,
allowing relative price positions, high--low ranges, trends, and geometric configurations to be viewed and processed jointly.
Two-dimensional convolutional neural networks (2D CNNs) can exploit this representation to learn hierarchical and nonlinear patterns
without requiring researchers to prespecify the relevant chart formations \citep{JiangKellyXiu2023}.

In this respect, applying a 2D CNN to candlestick charts provides a computational analogue to chart-based technical analysis.
Technical analysts inspect historical charts for recurring formations that may signal future price movements,
whereas the CNN learns return-predictive visual patterns systematically from data \citep{JiangKellyXiu2023,Radfar2025}.
\citet{JiangKellyXiu2023} show that CNNs applied to stock-chart images identify return-predictive price patterns and support profitable out-of-sample portfolio strategies,
while \citet{JeongByunKim2026} document related evidence in the Korean stock market.

Chart-based technical analysis,
however, commonly supplements candlestick charts with trading volume and manually constructed indicators intended to summarize trend, momentum, and volatility.
Trading volume is not constructed from OHLC prices and may convey information about the quality of market signals that
cannot be inferred from price statistics alone \citep{BlumeEasleyOHara1994},
whereas the trend, momentum, and volatility indicators considered here are constructed from historical OHLC data.

Although these indicators do not introduce an independent non-price data source,
their economically motivated transformations may still make particular features of price dynamics easier for a forecasting model to use.
It nevertheless remains unclear whether these OHLC-derived transformations provide incremental out-of-sample economic value once OHLC is included,
and whether the answer depends on data representation or model architecture.
These considerations give rise to two empirical questions.
First, do trading volume and the trend, momentum, and volatility indicator groups generate economically meaningful out-of-sample portfolio performance when used separately?
Second, do these technical-indicator groups provide incremental out-of-sample economic value once OHLC is included?

To address these questions,
we first evaluate the standalone economic value of trading volume and the trend, momentum, and volatility indicator groups.
We then test whether adding these technical-indicator groups to OHLC candlestick charts provides incremental out-of-sample economic value.
We also evaluate OHLC + Volume as a benchmark augmented specification because volume,
unlike the technical indicators, is not constructed from OHLC.

Using daily data for 1,707 stocks listed on the main boards of the Chinese A-share market from 2013 to 2025,
we construct approximately 3.36 million stock-window samples from 20-day market histories and predict the direction of cumulative returns over the subsequent five trading days.
For the primary image-based analysis,
we represent recent market histories as multi-channel images and estimate a 2D CNN.
We compare OHLC, trading volume, and the three technical-indicator groups both separately and in specifications that augment OHLC with additional inputs.
To determine whether the findings depend on data representation or model architecture,
we repeat the same comparisons by applying an $\ell_2$-regularized logistic regression, a 1D CNN, and an LSTM to numerical representations of the same samples.

We evaluate economic performance over the 2021--2025 out-of-sample period using equal-weighted decile portfolios with non-overlapping scheduled five-trading-day holding periods.
High-minus-Low (H--L) returns, Sharpe ratios, and Fama--French factor-adjusted alphas are accompanied by formal pairwise tests.
The analysis also considers traditional trading strategies, proportional transaction costs, and trading constraints specific to the Chinese A-share market.

The empirical results yield two central findings.
First, specifications based separately on OHLC, trading volume, and each technical-indicator group generate significantly positive H--L returns and factor-adjusted alphas,
together with economically meaningful Sharpe ratios; the OHLC-only specifications generally perform best under the nonlinear models.
Second, and central to the paper,
adding the technical-indicator groups to OHLC does not systematically improve H--L returns, Sharpe ratios, or factor-adjusted alphas across the four model architectures.
The only statistically significant improvements among the augmented specifications arise in the LSTM Sharpe ratios for OHLC + Volume and All features.
However, the All features specification does not significantly outperform OHLC + Volume in either H--L return or Sharpe ratio.
This pattern is consistent with the LSTM Sharpe-ratio improvement being associated primarily with trading volume rather than with the technical-indicator groups.
Additional benchmark and implementation analyses using the baseline OHLC-only 2D CNN show that it compares favorably with the four traditional trading strategies.
Both the H--L and High-only portfolios retain positive and statistically significant net returns at proportional one-way costs of up to 10 basis points;
the screened long-only portfolios also retain positive net returns that are generally statistically significant under a one-way transaction cost of 10 basis points.

This paper contributes to the existing literature in two ways.
It provides the first large-scale evidence from the Chinese A-share market on
the standalone and incremental economic value of the technical-indicator groups examined here.
It also evaluates these standalone and incremental comparisons across image and numerical representations
and across linear, convolutional, and recurrent model architectures.
The broadly consistent evidence indicates that the limited incremental value of these technical-indicator groups is not specific to
the candlestick-image representation or the 2D CNN architecture.

The remainder of the paper proceeds as follows.
Section~\ref{sec:data} describes the data, input representations, and sample construction.
Section~\ref{sec:models} presents the prediction models and estimation procedures.
Section~\ref{sec:portfolio_2DCNN} reports the portfolio results based on the 2D CNN,
and Section~\ref{sec:portfolio_numerical} presents the corresponding evidence from the numerical-input models.
Section~\ref{sec:implementation} examines transaction costs and trading constraints.
Section~\ref{sec:conclusion} concludes.

\section{Data preparation}
\label{sec:data}
\subsection{Data source and preprocessing}
This study employs a sample of 1,707 A-share stocks listed on the main boards of the Shanghai Stock Exchange (SSE) and the Shenzhen Stock Exchange (SZSE).
The sample comprises all stocks that were initially listed before January 1, 2011.
Although the sample is selected using the pre-2011 listing cutoff, the empirical analysis uses daily observations only from January 1, 2013, to December 31, 2025.
This requirement ensures that every sample stock had been listed for at least two years before the beginning of the sample period,
thereby mitigating potential distortions associated with unusually high price volatility during the period immediately following an initial public offering (IPO).
Daily open, high, low, close, and trading-volume (OHLCV) data are obtained from the Baostock database (\url{https://baostock.com}).
The final sample period comprises 3,157 trading days.

To account for corporate actions, such as dividend distributions and rights issues, we apply a return-based backward-adjustment scheme.
This method reconstructs historical price series by scaling prices to ensure the continuity of daily returns,
effectively neutralizing artificial price jumps caused by non-market factors.
Importantly, this adjustment does not assume dividend reinvestment.
Instead,
it corresponds to a hypothetical strategy in which an investor liquidates the position immediately prior to the ex-dividend (or ex-rights) date and
reinvests the full capital at the same price on the ex-date.
Under this assumption, dividend payouts do not reduce invested capital, and rights issues do not require additional capital injections.
As a result, the adjusted price series reflects pure price-based returns and provides a consistent measure of capital appreciation,
while abstracting from the actual cash flows associated with corporate actions.

To ensure temporal consistency, we remove observations corresponding to trading suspensions.
After removing suspension records, each stock is represented by a chronologically ordered sequence of active trading observations,
which is essential for the correct computation of rolling-window statistics and technical indicators.

Based on the cleaned and adjusted OHLCV data, we compute a set of commonly used traditional technical indicators using the Python library ta (\url{https://github.com/bukosabino/ta}).
These indicators are grouped into three categories: trend, momentum, and volatility.
Because these indicators require historical observations for their computation,
missing values may arise at the beginning of each stock's time series when the required lookback period is not yet available.

For trend indicators, we compute the simple moving average (SMA) of closing prices using 5-, 10-, and 20-day windows.
The resulting SMAs capture the underlying price trend at different horizons.
We also calculate the Average Directional Index (ADX),
together with the positive and negative directional indicators ($+\mathrm{DI}$ and $-\mathrm{DI}$), using a 10-day window.
ADX measures the strength of the prevailing trend regardless of its direction,
whereas $+\mathrm{DI}$ and $-\mathrm{DI}$ capture positive and negative directional movements, respectively.
In addition, we construct the Moving Average Convergence Divergence (MACD) indicator (fast = 8, slow = 20, signal = 6),
which captures changes in trend dynamics through the convergence and divergence of moving averages.

For momentum indicators, we compute the Relative Strength Index (RSI) with window lengths of 5, 10, and 20 days, which measures the speed and persistence of recent price movements.
We also compute the stochastic oscillator (STOCH) with a 10-day lookback window and a 3-day smoothing window,
which reflects the relative position of the closing price within its recent trading range.

For volatility indicators, we construct Bollinger Bands using a 10-day moving average and a standard deviation multiplier of 2,
which capture the dispersion of prices around the trend.
We also compute Average True Range (ATR)-based Bands with a 10-day window,
where the bands are centered around the 5-day moving average and defined as $\text{SMA}_5 \pm \text{ATR}$, capturing time-varying price volatility.

\subsection{Image representation}
\begin{figure}[H]
	\centering
	\includegraphics[width=0.60\linewidth]{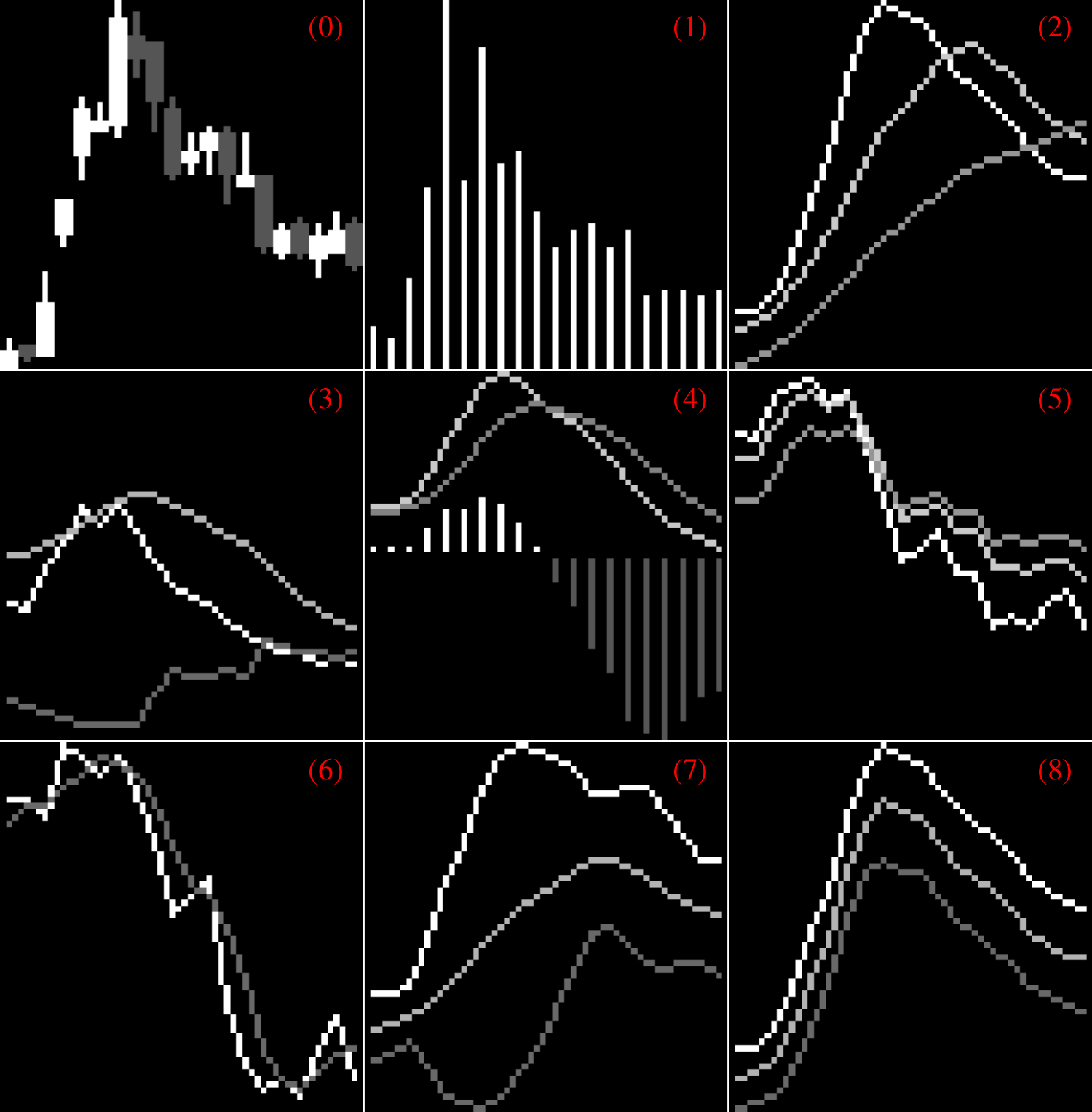}
	\caption{\textbf{Multi-channel image representation used in the 2D CNN model.}
	This figure presents the nine grayscale channels used as inputs to the 2D CNN.
	Channels (0) and (1) show the 20-day OHLC candlestick charts and trading volume bars, respectively;
	Channel (2) plots 5-, 10-, and 20-day moving averages of closing prices;
	Channel (3) presents ADX together with the positive and negative directional indicators;
	Channel (4) shows MACD;
	Channel (5) displays RSI with window lengths of 5, 10, and 20 days;
	Channel (6) presents STOCH;
	Channel (7) shows Bollinger Bands;
	and Channel (8) presents the ATR-based Bands.
	The channels collectively capture trend, momentum, and volatility information derived from daily stock data.}
	\label{fig:multi-imgs}
\end{figure}

To construct the inputs for the 2D CNN, we transform each 20-day window of OHLC prices, trading volume, and technical indicators into a structured nine-channel grayscale image.
Each image has dimensions of $61 \times 60 \times 9$ (height $\times$ width $\times$ channels), as illustrated in Figure~\ref{fig:multi-imgs}.
The horizontal axis represents time, with each of the 20 trading days occupying three consecutive pixels, resulting in a width of 60 pixels.
The vertical axis represents channel-specific scaled price, volume, or indicator values.
All channels use a black background (pixel value 0), with different grayscale intensities distinguishing graphical elements.
This representation retains the structural patterns in the input data while reducing storage and computational costs relative to RGB images.

The nine channels are designed to capture complementary aspects of market dynamics and are categorized into market data, trend, momentum, and volatility groups.
Channel (0) represents OHLC candlestick charts, while channel (1) captures trading volume.
Channels (2) through (4) capture trend information, including SMA, ADX, and MACD.
Channels (5) and (6) capture momentum via RSI and STOCH.
Channels (7) and (8) characterize volatility through Bollinger Bands and ATR-based Bands.

We apply channel-specific normalization within each input window.
Each of the candlestick, SMA, Bollinger Band, and ATR-based Band channels is independently scaled using the minimum and maximum values of the price series displayed in that channel.
Volume is normalized by its maximum value within the window, while RSI, STOCH, ADX, $+\mathrm{DI}$, and $-\mathrm{DI}$ are plotted on a fixed scale of 0 to 100.
For MACD, the difference line (DIF), signal line (DEA), and histogram series share a symmetric scale based on their maximum absolute value,
with zero positioned at the vertical center of the image.

The image generation procedure utilizes custom rendering functions that operate directly on pixel arrays.
Candlestick charts are rendered using high-low vertical lines and rectangular bodies, with grayscale intensity encoding price direction.
Trading volume is visualized as vertical bars normalized by the peak volume within the window.
Continuous indicator curves are rasterized using Bresenham's line algorithm to connect consecutive observations.

For the MACD indicator, the DIF and DEA lines are drawn as continuous curves,
while the histogram is represented by vertical bars extending above or below the zero line.
For multi-component indicators, such as the K and D lines in STOCH, varying grayscale intensities are used to distinguish between components.
This deterministic construction approach ensures reproducibility of the data construction process.

\subsection{Scaling the data underlying the images}
To construct numerical inputs for the logistic regression, 1D CNN, and LSTM models, we scale the data underlying the image representations.
These data differ in scale across variables and in magnitude across stocks and time windows.
We therefore apply window-specific linear scaling to make the numerical inputs comparable and prevent the model results from being driven by these differences.
Using the same underlying variables also places the numerical models and the image-based 2D CNN on a comparable informational basis.

Except for MACD, all variables are linearly scaled to $[0,1]$ within each 20-day input window.
OHLC prices are jointly scaled using the minimum low and maximum high within the window.
Trading volume is divided by its window maximum, while the 5-, 10-, and 20-day SMAs are jointly scaled using their overall minimum and maximum.
ADX and its directional components, RSI, and STOCH are divided by 100.

Because MACD can take both positive and negative values,
its DIF, DEA, and histogram series are jointly scaled to $[-1,1]$ using their largest absolute value within the window, with zero remaining unchanged.
The Bollinger and ATR-based Bands are represented by their scaled middle lines and half-widths.
For each indicator, the middle line is scaled using the minimum lower band and maximum upper band, and the half-width is divided by the same range.
All window-specific scaling parameters are calculated solely from observations within the corresponding input window, thereby avoiding look-ahead bias.

The resulting numerical data represent each sample as a $20\times23$ matrix,
with rows corresponding to trading days and columns corresponding to individual features.
The 23 features comprise four OHLC prices, trading volume, three SMAs, ADX and its two directional components, the MACD DIF, DEA, and histogram series,
three RSI measures, the STOCH K and D lines, the Bollinger Band middle line and half-width, and the ATR-based Band middle line and half-width.
This differs from the nine-channel image representation because a single image channel may contain multiple series, whereas each series is stored as a separate numerical feature. 

\subsection{Sample construction and prediction setup}
We construct prediction samples from rolling 20-trading-day windows of stock-level data.
For expositional clarity, the following discussion focuses primarily on the construction of image samples.
The numerical samples used by the logistic regression, 1D CNN, and LSTM models
are constructed from exactly the same stock-window observations and follow the same labeling, filtering, and temporal-splitting procedures.
The only difference lies in the input representation:
each window is converted either into a multi-channel image for the 2D CNN or into the scaled numerical matrix described in the preceding subsection.
The image and numerical inputs are therefore alternative representations of the same underlying observations rather than separate samples.

Following the daily rolling-window approach of \citet{JeongByunKim2026},
the training and validation samples are generated by advancing the 20-day input window by one trading day at a time.
Unlike their random training--validation split, we assign samples chronologically to separate training and validation periods.
This procedure substantially increases the number of examples available for model estimation and validation and allows the models to learn from every eligible window-ending date.
Adjacent windows consequently share 19 of their 20 observations, and their five-trading-day prediction intervals may also overlap.
These samples are therefore not interpreted as statistically independent observations;
rather, the overlapping construction is used to provide a dense set of input patterns for model training.

For out-of-sample evaluation,
window-ending dates in the test sample advance in increments of five market trading days.
This design aligns the out-of-sample prediction frequency with the portfolio strategy:
stocks are sorted on each prediction date,
and the resulting portfolios are held for five market trading days before rebalancing.
Consequently, adjacent scheduled portfolio holding periods do not overlap.

Candidate windows are retained only when they contain 20 valid trading observations,
do not include any day on which the stock is classified as Special Treatment (ST), and have complete values for all required input variables.
Observations corresponding to trading suspensions are removed before window construction.
In the market-calendar-aligned test set, a stock with insufficient trading observations within a designated window is therefore excluded from that window.
These filters are applied at the stock-window level, so an invalid window does not require the removal of other valid observations for the same stock.

For each valid window ending on trading day $t$,
the input contains information from the preceding 20 active trading observations, including day $t$.
The prediction target is the cumulative stock return from trading day $t$ to the fifth subsequent active trading day $t+5$.
The binary label equals one if the adjusted closing price on day $t+5$ exceeds that on day $t$, and zero otherwise.
Thus, all input variables are observed no later than day $t$, whereas the label depends only on the subsequent return, preventing look-ahead bias.

In addition to the image or numerical input and its binary label, we retain the window-ending date, stock code, and realized five-trading-day return.
These variables are used to align predictions across models and construct the decile portfolios in the out-of-sample analysis.

Samples are assigned to three calendar periods: 2013--2018 for training, 2019--2020 for validation, and 2021--2025 for testing.
This procedure produces approximately 3.36 million stock-window observations.
Because adjacent windows overlap within both the training and validation sets,
their counts refer to model inputs rather than statistically independent observations.
Table~\ref{tab:dataset_split} summarizes the temporal split and the proportion of positive five-trading-day returns in each subset.
The proportions are all close to 50\%, indicating that the three subsets are well-balanced for binary classification.
\begin{table}[H]
    \centering
    \caption{\textbf{Summary of dataset split.}}
    \label{tab:dataset_split}
    \begin{tabular*}{\textwidth}{@{\extracolsep{\fill}} lccc}
        \toprule
        Dataset        & Period     & Sample count   & Positive return ratio \\
        \midrule
        Training set   & 2013--2018 & 2,219,123      & 50.5\% \\
        Validation set & 2019--2020 &   751,829      & 49.7\% \\
        Test set       & 2021--2025 &   388,914      & 48.4\% \\
        \bottomrule
    \end{tabular*}
\end{table}

\section{Prediction models and estimation procedures}
\label{sec:models}
This section presents the 2D CNN, logistic regression, 1D CNN, and LSTM models and describes their estimation procedures.

\subsection{Architecture of the 2D CNN}
The model takes as input multi-channel images that encode historical market data, including candlestick charts, trading volume, and the technical indicators examined here.
Together, these channels represent different aspects of recent market dynamics,
enabling the 2D CNN to learn jointly from price, volume, trend, momentum, and volatility patterns.
Formally, each observation is transformed into a tensor $\mathbf{X} \in \mathbb{R}^{C \times H \times W}$,
where $C$ represents the number of channels, and $H$ and $W$ denote the spatial dimensions (height and width).

\begin{figure}[H]
	\centering
	\includegraphics[width=0.85\linewidth]{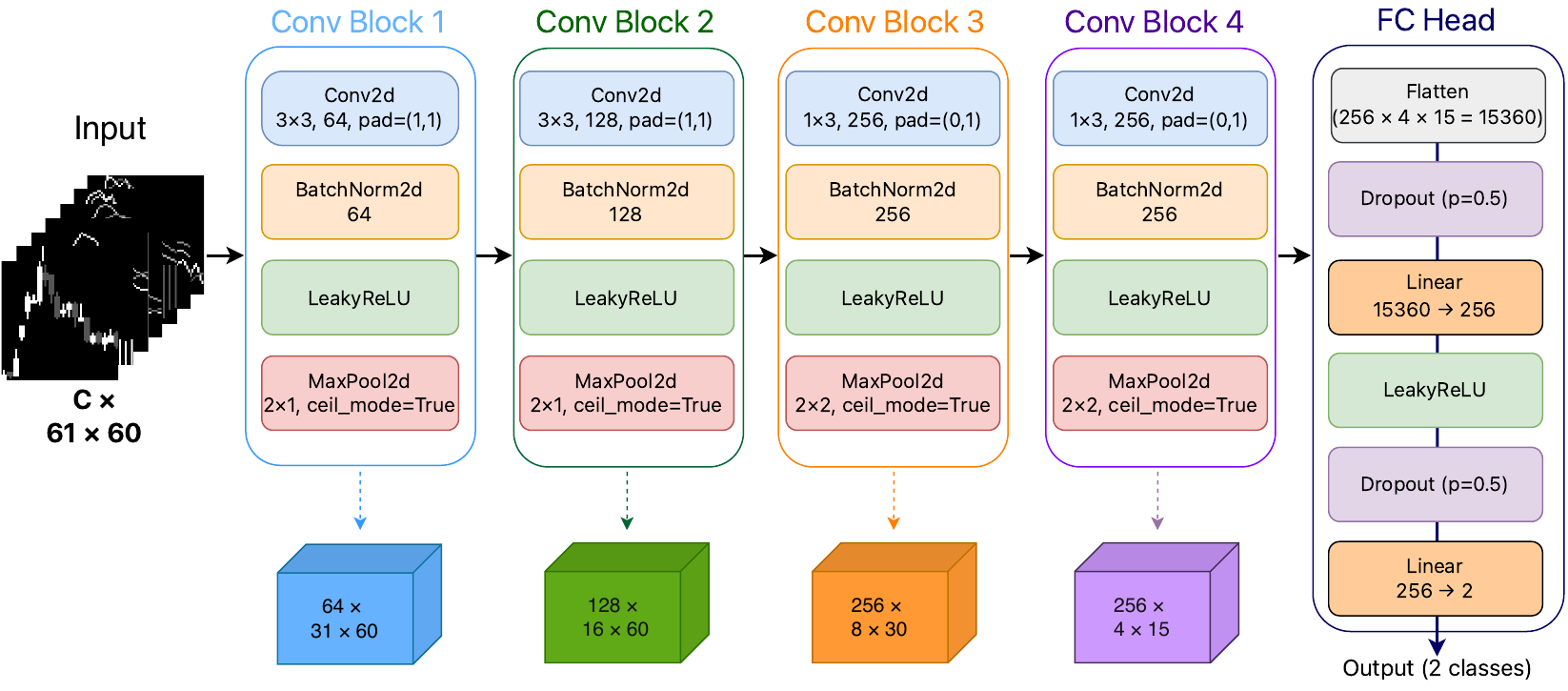}
	\caption{\textbf{Architecture of the proposed 2D CNN.}
	The model takes as input multi-channel images of size $C \times 61 \times 60$, where $C$ denotes the number of input channels.
	It consists of four convolutional blocks, each comprising a convolutional layer, batch normalization, Leaky ReLU activation, and max-pooling.
	The first two blocks employ asymmetric pooling with a kernel size of $2 \times 1$,
	which downsamples the feature maps along the vertical dimension while preserving the horizontal dimension.
	The last two blocks use standard $2 \times 2$ pooling to downsample both spatial dimensions.
	The extracted features are flattened and fed into an FC head with dropout regularization, producing a two-class output.}
	\label{fig:2DCNNArchitecture}
\end{figure}

The network architecture, illustrated in Figure~\ref{fig:2DCNNArchitecture},
adopts a deep 2D CNN structure composed of a sequence of convolutional blocks followed by a fully connected (FC) head.
Each convolutional block comprises
a convolutional layer,
batch normalization,
a leaky rectified linear unit (Leaky ReLU) activation function,
and a max-pooling operation.
Convolutional layers extract local spatial patterns,
batch normalization stabilizes the training process,
and the Leaky ReLU activation introduces nonlinearity to enhance model flexibility.
Max-pooling reduces spatial dimensionality and helps mitigate overfitting.

Following the convolutional backbone, the extracted feature maps are flattened and passed through the FC head.
Dropout regularization is applied to the dense layers to mitigate overfitting.
The final layer outputs two logits corresponding to the binary classification task of predicting positive versus non-positive future returns.

\subsection{Logistic regression specification}
As a linear benchmark, we employ an $\ell_2$-regularized logistic regression using the scaled numerical data.
The model uses observations from the entire 20-day input window described in Section~\ref{sec:data}.
For each sample, the daily variables included in the input are concatenated in chronological order to form a single predictor vector.
The MACD histogram is excluded because it is linearly determined by DIF and DEA.

Following the window-specific scaling procedure, the predictors are standardized using means and standard deviations estimated from the training set.
The same transformation is subsequently applied to the validation and test sets.
The inverse regularization parameter is selected by minimizing log loss on the validation set.
The fitted model produces the predicted probability of a positive subsequent five-trading-day return.

\subsection{1D CNN and LSTM architectures}
To examine whether the findings extend beyond the image representation and 2D CNN architecture, we additionally employ 1D CNN and LSTM models using the scaled numerical data.
Each numerical sample is organized as a $20\times K$ matrix,
where the rows represent trading days and $K$ denotes the number of variables in the feature groups included in a given specification.
The feature groups correspond to those used in the image-based analysis:
the OHLC group contains the four price variables;
volume contains trading volume;
trend contains the SMA, ADX, and MACD series;
momentum contains the RSI and STOCH series;
and volatility contains the middle lines and half-widths of the Bollinger and ATR-based Bands.
We also consider specifications combining OHLC with each of the other feature groups and a specification containing all 23 numerical features.
Unlike the nine-channel images, in which one channel may display multiple series, each series constitutes a separate feature in the numerical inputs.

The 1D CNN treats the numerical variables as input channels and applies convolution along the 20-day temporal dimension.
To facilitate comparison with the 2D CNN, it follows the same four-block structure and channel progression of 64, 128, 256, and 256.
Each block consists of a one-dimensional convolution with a kernel size of 3, batch normalization, Leaky ReLU activation, and max-pooling with a kernel size of 2.
The successive pooling operations reduce the temporal dimension from 20 to 2.
The resulting features are flattened and passed through a classification head consisting of a dropout layer,
a fully connected layer with 256 units, Leaky ReLU activation, a second dropout layer, and a final two-unit output layer.

The LSTM processes the numerical input directly as a sequence of 20 daily observations.
It consists of a unidirectional LSTM layer with a hidden dimension of 256.
The output of the LSTM layer on the last day of the input window is used as the sequence representation.
This output is layer-normalized and passed through a classification head comprising a dropout layer,
a fully connected layer with 256 units, Leaky ReLU activation, a second dropout layer, and a final two-unit output layer.
Both the 1D CNN and LSTM therefore produce two logits corresponding to positive and non-positive subsequent five-trading-day returns.

\subsection{Neural-network training procedure}
We formulate the return prediction task as a binary classification problem,
where the label equals one if the subsequent five-trading-day cumulative return is positive and zero otherwise.
Each neural network produces two logits, which serve as inputs to the cross-entropy loss function.

We initialize the convolutional and fully connected layers using the Xavier uniform distribution \citep{Xavier2010} and employ the AdamW algorithm \citep{AdamW2019} for optimization,
which decouples weight decay from gradient updates and helps control model complexity.
To stabilize LSTM training, we clip the gradient norm at 1.0.
Training is performed using mini-batch stochastic optimization.
To facilitate convergence, we implement a cosine annealing learning rate schedule \citep{Cosine2017}.
This approach allows for relatively larger parameter updates in the early stages of training and finer adjustments as training progresses.
To reduce overfitting, we employ an early-stopping strategy based on validation set performance.
After each epoch, the model is evaluated on the validation set;
if validation accuracy fails to improve for a predefined number of consecutive epochs, training is terminated,
and the parameters corresponding to the best validation performance are restored.
Key training configurations and hyperparameters are summarized in Table~\ref{tab:training_params}.
For each neural-network architecture, the training configuration is held fixed across all input specifications.

Given the stochastic nature of deep neural network training, we perform multiple independent repetitions with different random initializations.
For each repetition, we retain the model parameters that achieve the highest validation accuracy.
During the prediction phase,
the final predicted probability is obtained by averaging the positive-class softmax probabilities across the five trained models.
This ensemble approach reduces estimation variance and enhances the robustness of our predictions.
\begin{table}[H]
	\centering
	\caption{\textbf{Training configurations and hyperparameters.}}
	\label{tab:training_params}
	\begin{tabular*}{0.7\textwidth}{@{\extracolsep{\fill}}lc}
		\toprule
		Configuration / hyperparameter        & Value \\
		\midrule
		Batch size                            & 128 \\
		Maximum epochs                        & 50 \\
		Independent training repetitions      & 5 \\
		Optimizer                             & AdamW \\
		Learning rate scheduler               & Cosine annealing \\
		Cosine annealing horizon              & 100 epochs\\
		Initial learning rate                 & $5 \times 10^{-4}$ \\
		Lower learning rate bound             & $10^{-6}$ \\
		Weight decay                          & $10^{-2}$ \\
		Gradient clipping threshold           & 1.0 (LSTM only) \\
		Dropout rate                          & 0.5\\
		Early stopping patience               & 5 epochs \\
		\bottomrule
	\end{tabular*}
\end{table}

\section{Portfolio performance based on 2D CNN predictions}
\label{sec:portfolio_2DCNN}
\subsection{Decile portfolio performance}
We evaluate both the cross-sectional ranking ability and the economic value of the 2D CNN forecasts using equal-weighted decile portfolios.
As described in Section~\ref{sec:data}, out-of-sample portfolio formation dates are spaced five market trading days apart.
At each formation date, we average the out-of-sample predicted probabilities of a positive five-trading-day return across five independently trained models.
Stocks are then sorted by the averaged probability into ten portfolios, from Low to High, and held for five trading days until the next rebalancing.
Consequently, adjacent scheduled portfolio holding periods do not overlap.
This construction allows us to examine whether realized returns increase with the model-implied ranking.
We then summarize the return spread associated with this ranking using the H--L portfolio, which takes a long position in High and a short position in Low.

Table~\ref{tab:2dcnn} reports annualized mean returns (Ret), annualized Sharpe ratios (SR), and average monthly turnover \citep{GuKellyXiu2020,JiangKellyXiu2023}.
Panel A presents models based on individual feature groups, whereas Panel B presents models that combine OHLC with additional feature groups.
The row labeled $\Delta$ reports the H--L performance of the OHLC-only model minus that of each alternative specification,
so a positive value indicates that the OHLC-only model performs better.
For all portfolio-performance analyses in this study,
the significance of mean H--L returns and pairwise differences in mean H--L returns is assessed using two-sided tests with Newey--West standard errors \citep{NeweyWest1987},
whereas pairwise differences in Sharpe ratios are evaluated using two-sided HAC delta-method tests \citep{Lo2002}.
All HAC-based tests use one lag.
\begin{table}[H]
	\centering
	\caption{\textbf{Decile portfolio performance based on 2D CNN forecasts.}
	This table reports the out-of-sample performance of equal-weighted decile portfolios across alternative input specifications.
	Panel A presents individual feature groups,
	whereas Panel B presents OHLC combined with additional inputs,
	including an All features specification containing every input group.
	Ret and SR denote the annualized mean return and annualized Sharpe ratio, respectively,
	while Turnover denotes the average monthly turnover of the H--L portfolio.
	The $\Delta$ row reports the corresponding performance differences relative to the OHLC-only specification.
	***, **, and * denote significance at the 1\%, 5\%, and 10\% levels, respectively.}
	\label{tab:2dcnn}
	\begin{tabular*}{\textwidth}{@{\extracolsep{\fill}} l *{5}{cc}}
		\toprule
		\multicolumn{11}{l}{Panel A: Individual feature groups}\\
		\midrule
		&\multicolumn{2}{c}{OHLC}
		&\multicolumn{2}{c}{Volume}
		&\multicolumn{2}{c}{Trend}
		&\multicolumn{2}{c}{Momentum}
		&\multicolumn{2}{c}{Volatility} \\
		         \cmidrule(lr){2-3}            \cmidrule(lr){4-5}             \cmidrule(lr){6-7}             \cmidrule(lr){8-9}             \cmidrule(lr){10-11}
		         & Ret     & SR                & Ret     & SR                 & Ret     & SR                 & Ret     & SR                  & Ret     & SR \\
		\midrule
		Low      &-0.06    &-0.29              &-0.02    &-0.07               & 0.01    & 0.04               & 0.04    & 0.21                & 0.08    & 0.37  \\
		2        & 0.05    & 0.25              & 0.16    & 0.73               & 0.10    & 0.47               & 0.07    & 0.32                & 0.07    & 0.32  \\
		3        & 0.08    & 0.39              & 0.18    & 0.85               & 0.14    & 0.67               & 0.09    & 0.43                & 0.12    & 0.61  \\
		4        & 0.13    & 0.62              & 0.15    & 0.73               & 0.13    & 0.62               & 0.12    & 0.58                & 0.14    & 0.70  \\
		5        & 0.19    & 0.92              & 0.16    & 0.79               & 0.14    & 0.68               & 0.12    & 0.58                & 0.17    & 0.82  \\
		6        & 0.20    & 1.00              & 0.19    & 0.97               & 0.18    & 0.91               & 0.18    & 0.89                & 0.17    & 0.85  \\
		7        & 0.21    & 1.01              & 0.21    & 1.05               & 0.19    & 0.89               & 0.20    & 0.95                & 0.19    & 0.90  \\
		8        & 0.24    & 1.18              & 0.20    & 0.99               & 0.22    & 1.03               & 0.26    & 1.22                & 0.22    & 1.02  \\
		9        & 0.27    & 1.29              & 0.20    & 0.98               & 0.24    & 1.12               & 0.25    & 1.16                & 0.22    & 1.03  \\
		High     & 0.31    & 1.40              & 0.18    & 0.88               & 0.28    & 1.24               & 0.30    & 1.32                & 0.24    & 1.08  \\
		H--L     & 0.38*** & 3.37              & 0.20*** & 1.54               & 0.27*** & 2.29               & 0.26*** & 2.22                & 0.17*** & 1.43  \\
		$\Delta$ & --      & --                & 0.18*** & 1.83***            & 0.11**  & 1.08**             & 0.12*** & 1.15***             & 0.21*** & 1.95***\\
		Turnover & \multicolumn{2}{c}{676\%}   & \multicolumn{2}{c}{615\%}    & \multicolumn{2}{c}{653\%}    & \multicolumn{2}{c}{670\%}     & \multicolumn{2}{c}{681\%} \\
		
		\midrule
		\multicolumn{11}{l}{Panel B: Combined feature groups}\\
		\midrule
		&\multicolumn{2}{c}{OHLC + Volume}
		&\multicolumn{2}{c}{OHLC + Trend}
		&\multicolumn{2}{c}{OHLC + Momentum}
		&\multicolumn{2}{c}{OHLC + Volatility}
		&\multicolumn{2}{c}{All features} \\
		        \cmidrule(lr){2-3}            \cmidrule(lr){4-5}            \cmidrule(lr){6-7}            \cmidrule(lr){8-9}            \cmidrule(lr){10-11}
		        & Ret     & SR                & Ret     & SR                & Ret     & SR                & Ret     & SR                 & Ret     & SR \\
		\midrule
		Low     &-0.10    &-0.45              &-0.03    &-0.14              &-0.02    &-0.11              &-0.05    &-0.22               &-0.06    &-0.29  \\
		2       & 0.06    & 0.29              & 0.06    & 0.31              & 0.05    & 0.26              & 0.05    & 0.23               & 0.06    & 0.26  \\
		3       & 0.14    & 0.64              & 0.08    & 0.38              & 0.09    & 0.44              & 0.10    & 0.48               & 0.09    & 0.44  \\
		4       & 0.13    & 0.65              & 0.13    & 0.65              & 0.12    & 0.59              & 0.13    & 0.64               & 0.11    & 0.55  \\
		5       & 0.19    & 0.91              & 0.15    & 0.76              & 0.15    & 0.75              & 0.18    & 0.88               & 0.18    & 0.86  \\
		6       & 0.20    & 0.98              & 0.19    & 0.92              & 0.19    & 0.92              & 0.21    & 1.02               & 0.21    & 1.02  \\
		7       & 0.21    & 1.01              & 0.20    & 0.98              & 0.22    & 1.05              & 0.21    & 1.07               & 0.22    & 1.08  \\
		8       & 0.25    & 1.22              & 0.28    & 1.32              & 0.24    & 1.14              & 0.22    & 1.05               & 0.25    & 1.17  \\
		9       & 0.26    & 1.27              & 0.26    & 1.23              & 0.29    & 1.32              & 0.27    & 1.26               & 0.26    & 1.26  \\
		High    & 0.29    & 1.34              & 0.30    & 1.30              & 0.30    & 1.34              & 0.30    & 1.33               & 0.31    & 1.42  \\
		H--L    & 0.39*** & 3.54              & 0.33*** & 2.96              & 0.32*** & 2.81              & 0.35*** & 2.94               & 0.38*** & 3.63  \\
		$\Delta$&-0.02    &-0.17              & 0.05*   & 0.41              & 0.06*   & 0.56*             & 0.03    & 0.43*              &0.00     &-0.26  \\
		Turnover& \multicolumn{2}{c}{666\%}   & \multicolumn{2}{c}{668\%}   & \multicolumn{2}{c}{667\%}   & \multicolumn{2}{c}{670\%}    & \multicolumn{2}{c}{668\%} \\
		\bottomrule
	\end{tabular*}
\end{table}

Across both panels, the realized portfolio returns generally increase from Low to High, although the intermediate deciles are not perfectly monotonic in some specifications.
Most notably, the OHLC-only specification produces an annualized return of $-0.06$ for Low and $0.31$ for High.
For every other specification, High also earns a higher return than Low.
This broad ordering indicates that the 2D CNN forecasts contain meaningful information for ranking stocks cross-sectionally.

Consistent with this sorting pattern, Panel A shows that the 2D CNN produces economically and statistically significant return spreads from each individual feature group.
The OHLC-only model delivers an annualized H--L return of 0.38 and a Sharpe ratio of 3.37, with the mean return significant at the 1\% level.
Models based separately on Volume, Trend, Momentum, and Volatility also generate significantly positive H--L returns of 0.20, 0.27, 0.26, and 0.17, respectively.
These results indicate that each feature group contains standalone information useful for ranking future stock returns.
Nevertheless, the OHLC-only model performs substantially better than every other individual-group specification.
The differences in both H--L returns and Sharpe ratios are statistically significant in all four comparisons.
Thus, although the manually constructed technical indicators possess predictive content on their own, their standalone performance is weaker than that of the OHLC representation.

Panel B directly evaluates whether additional feature groups improve upon the OHLC-only model.
Adding Volume raises the annualized H--L return from 0.38 to 0.39 and the Sharpe ratio from 3.37 to 3.54, but neither difference is statistically significant.
Adding Trend or Momentum instead reduces the H--L return to 0.33 and 0.32, respectively; both return differences are significant at the 10\% level.
The OHLC + Momentum specification also has a significantly lower Sharpe ratio at the 10\% level.
Adding Volatility produces an H--L return of 0.35 and a Sharpe ratio of 2.94.
Its return difference relative to OHLC is not statistically significant, although its Sharpe ratio is significantly lower at the 10\% level.
Finally, the All features model generates an H--L return of 0.38 and a Sharpe ratio of 3.63; neither differs significantly from the corresponding OHLC-only result.

Monthly turnover is uniformly high, ranging from 615\% to 681\%, but varies relatively little across specifications.
This similarity suggests that differences in gross portfolio performance are unlikely to be driven primarily by differences in trading intensity.
Nevertheless, the absolute turnover levels imply substantial rebalancing and potentially material transaction costs, which may reduce the implementable returns of these strategies.
The implications of high turnover and trading costs for strategy implementation are discussed in a later section.

Taken together,
the OHLC-only specification exhibits a stronger relation between model rankings and realized returns than the specifications based separately on
Volume, Trend, Momentum, or Volatility,
as reflected in its significantly higher H--L return and Sharpe ratio.
This evidence is consistent with the 2D CNN learning more economically useful return-predictive patterns from OHLC images than from the standalone alternative inputs.
At the same time, augmenting OHLC with the additional feature groups yields no statistically detectable improvement in H--L returns or Sharpe ratios.

\subsection{Cumulative performance over time}
\begin{figure}[H]
	\centering
	\includegraphics[width=0.60\textwidth]{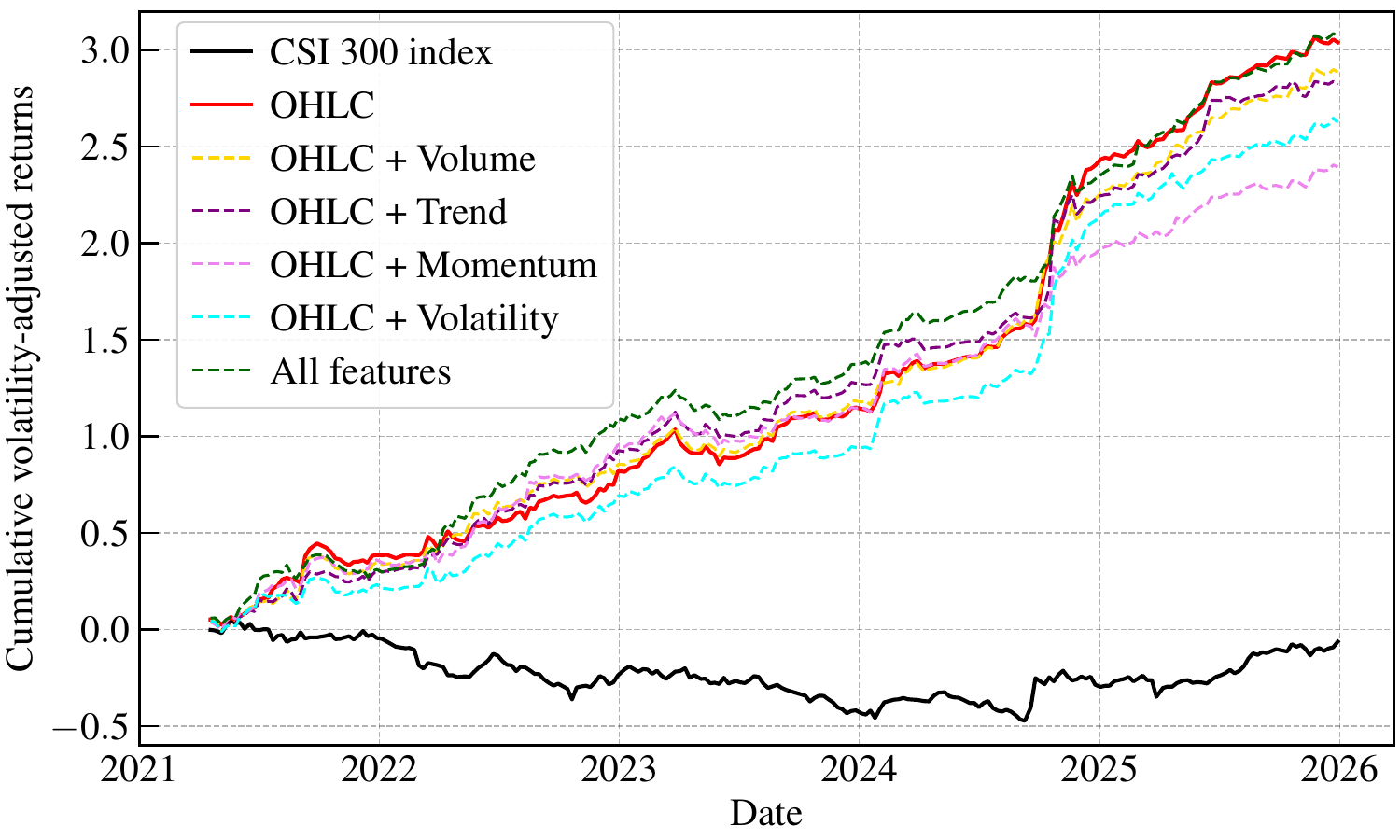}
	\caption{\textbf{Cumulative volatility-adjusted log returns of H--L portfolios based on 2D CNN predictions.}
	This figure plots the cumulative log returns of volatility-adjusted H--L portfolios constructed from 2D CNN predictions
	under alternative input configurations and the cumulative log return of the CSI 300 Index, which serves as the benchmark for volatility adjustment.
	The red solid curve represents the OHLC-only model, while the black solid curve represents the CSI 300 Index.
	The colored dashed curves represent models that augment OHLC with trading volume or technical-indicator groups and the model using all features.}
	\label{fig:CumRet_2DCNN}
\end{figure}

Figure~\ref{fig:CumRet_2DCNN} plots the cumulative volatility-adjusted log returns of the H--L portfolios generated by the OHLC-only and augmented-input 2D CNN specifications,
together with the cumulative log return of the CSI 300 Index over the corresponding five-market-day evaluation periods.
Each H--L portfolio return is scaled by the ratio of the rolling volatility of the corresponding five-market-day CSI 300 return series to that of the strategy,
with both volatility estimates lagged by one portfolio formation period so that the scaling relies only on information available before the corresponding portfolio return is realized.
The adjusted portfolio returns and the corresponding CSI 300 returns are then transformed into log returns and cumulatively summed.

The OHLC-containing specifications exhibit broadly similar movements over time, but their cumulative return paths are not uniformly close throughout the test period.
Their relative rankings change over time, and no augmented-input specification consistently dominates the OHLC-only model.
By the end of the sample, the cumulative volatility-adjusted log returns of the OHLC-only and All features specifications are 3.04 and 3.07, respectively.
The corresponding values are 2.89 for OHLC + Volume, 2.82 for OHLC + Trend, 2.39 for OHLC + Momentum, and 2.63 for OHLC + Volatility.
Thus, the All features specification ends the sample close to the OHLC-only model,
whereas several augmented specifications increasingly lag behind it during the later part of the test period.

This time-series evidence complements the portfolio results in Table~\ref{tab:2dcnn}.
Although the augmented specifications outperform the OHLC-only model during some subperiods, none delivers a consistently larger cumulative return throughout the full test period.
This finding is consistent with the absence of a statistically significant improvement in H--L return or Sharpe ratio.
The terminal ranking of the cumulative curves need not coincide exactly with the rankings
based on annualized returns and Sharpe ratios because the curves incorporate time-varying volatility scaling and log compounding.
The model-based portfolios also accumulate substantially larger returns than the CSI 300 Index over the test period,
providing further evidence that the cross-sectional rankings generated by the 2D CNN contain economically useful out-of-sample predictive information.

\subsection{Comparison with traditional trading strategies}
The preceding results compare alternative inputs within the same 2D CNN framework.
We next assess the economic value of the OHLC-based forecasts relative to four conventional price-based trading signals:
12--2 momentum (MOM), short-term reversal (STR), weekly short-term reversal (WSTR), and trend (TREND).
Let $P_t$ denote a stock's adjusted closing price on portfolio formation date $t$.
MOM is defined as $P_{t-20}/P_{t-250}-1$ and therefore measures price appreciation from approximately one year to one month before portfolio formation \citep{JegadeeshTitman1993}.
STR and WSTR are defined as $-(P_t/P_{t-20}-1)$ and $-(P_t/P_{t-5}-1)$, respectively \citep{Jegadeesh1990, Lehmann1990}.
The reversal signals are signed so that higher values identify stocks with lower recent returns.

Following \citet{HanZhouZhu2016}, TREND combines normalized moving averages measured over 3, 5, 10, 20, 50, 100, 200, 400, 600, 800, and 1,000 trading days.
At each month-end, realized monthly returns are regressed cross-sectionally on the preceding month-end moving-average-to-price ratios.
The regression coefficients are averaged over the most recent 12 months,
and the latest available averaged coefficients are applied to the current normalized moving averages to generate the TREND signal.
This procedure updates the regression coefficients monthly and the TREND signal daily,
while ensuring that each portfolio formation date uses only coefficients available by that date.
For comparability with the 2D CNN portfolios,
all four signals are evaluated on the same out-of-sample formation dates using the same equal-weighted decile construction and five-trading-day holding period.

Table~\ref{tab:traditional_performance} reports the resulting portfolio performance.
At each portfolio formation date, stocks are sorted into deciles according to each signal.
The decile patterns are considerably less orderly than those produced by the OHLC-only 2D CNN.
MOM does not generate a positive winner-minus-loser spread during the test period:
High earns 0.06 annually, compared with 0.11 for Low, resulting in an insignificant H--L return of $-0.06$ and a Sharpe ratio of $-0.29$.
By contrast, STR, WSTR, and TREND produce significantly positive H--L returns of 0.31, 0.23, and 0.25, with Sharpe ratios of 1.36, 1.03, and 1.22, respectively.
However, portfolios formed by sorting stocks on the traditional signals do not exhibit a monotonic pattern in subsequent realized returns,
indicating limited cross-sectional ranking ability.
In contrast,
sorting stocks by the out-of-sample predicted probabilities generated by the OHLC-only 2D CNN produces a monotonic increase in realized returns from Low to High,
providing evidence of the model's stronger cross-sectional ranking ability.

The $\Delta$ row formally compares each traditional strategy with the OHLC-only 2D CNN, whose H--L portfolio earns 0.38 annually with a Sharpe ratio of 3.37.
The $\Delta$ values are calculated as the corresponding value for the OHLC-only specification minus that for the indicated strategy.
A positive value therefore indicates better performance by the OHLC-only 2D CNN.
Relative to MOM, the OHLC-only model raises the H--L return by 0.44 and the Sharpe ratio by 3.66, with both differences significant at the 1\% level.
Its return advantages over STR, WSTR, and TREND are 0.06, 0.15, and 0.12, respectively.
The return advantage is significant at the 10\% level relative to WSTR but is not statistically significant relative to STR or TREND.
In contrast, the Sharpe-ratio improvements of 2.02, 2.34, and 2.15 relative to these three strategies are all significant at the 1\% level.
Thus, the OHLC-based 2D CNN exhibits uniformly stronger risk-adjusted performance,
although the statistical significance of its mean-return advantage depends on the traditional benchmark considered.

Average monthly turnover ranges from 154\% for MOM to 685\% for WSTR, compared with 676\% for the OHLC-only 2D CNN.
MOM, STR, and TREND therefore require materially less trading than the CNN strategy, whereas WSTR has similar turnover.
Because the reported returns are before transaction costs, the lower turnover of some traditional strategies may narrow their net-performance disadvantage.

Overall, the reversal and TREND signals generate economically meaningful portfolio performance, whereas MOM performs poorly over this sample and forecast horizon.
The OHLC-only 2D CNN nevertheless delivers substantially stronger risk-adjusted performance than all four conventional rules and a more orderly cross-sectional return ranking.
These comparisons establish the relative portfolio performance of the competing signals,
but they do not determine whether the return spreads are explained by exposures to common equity factors.
We therefore examine factor-adjusted performance next.
\begin{table}[H]
	\centering
	\caption{\textbf{Decile portfolio performance of traditional trading strategies and comparison with the OHLC-only 2D CNN.}
	This table reports the out-of-sample performance of equal-weighted decile portfolios formed using the MOM, STR, WSTR, and TREND signals.
	Ret and SR denote the annualized mean return and annualized Sharpe ratio, respectively,
	while Turnover denotes the average monthly turnover of the H--L portfolio.
	The $\Delta$ row reports the corresponding performance differences relative to the OHLC-only 2D CNN specification.
	***, **, and * denote significance at the 1\%, 5\%, and 10\% levels, respectively.}
	\label{tab:traditional_performance}
	\begin{tabular*}{\textwidth}{@{\extracolsep{\fill}} l *{4}{cc}}
		\toprule
		        &\multicolumn{2}{c}{MOM}
				&\multicolumn{2}{c}{STR}
				&\multicolumn{2}{c}{WSTR}
				&\multicolumn{2}{c}{TREND}\\
		        \cmidrule(lr){2-3}              \cmidrule(lr){4-5}             \cmidrule(lr){6-7}              \cmidrule(lr){8-9}
		        & Ret     & SR                    & Ret     & SR                 & Ret     & SR                  & Ret     & SR \\
		\midrule
		Low     & 0.11    & 0.46                  &-0.15    &-0.57               &-0.12    &-0.52                &-0.07    &-0.28 \\
		2       & 0.15    & 0.69                  & 0.14    & 0.66               & 0.15    & 0.75                & 0.14    & 0.66 \\
		3       & 0.18    & 0.86                  & 0.21    & 1.05               & 0.19    & 0.96                & 0.17    & 0.84 \\
		4       & 0.20    & 0.97                  & 0.19    & 0.95               & 0.20    & 1.07                & 0.18    & 0.92 \\
		5       & 0.19    & 0.95                  & 0.23    & 1.15               & 0.23    & 1.19                & 0.22    & 1.09 \\
		6       & 0.24    & 1.16                  & 0.22    & 1.11               & 0.21    & 1.02                & 0.22    & 1.07 \\
		7       & 0.19    & 0.94                  & 0.21    & 1.01               & 0.25    & 1.20                & 0.19    & 0.95 \\
		8       & 0.17    & 0.80                  & 0.22    & 1.00               & 0.21    & 0.92                & 0.21    & 0.97 \\
		9       & 0.14    & 0.64                  & 0.20    & 0.84               & 0.21    & 0.85                & 0.18    & 0.78 \\
		High    & 0.06    & 0.23                  & 0.17    & 0.62               & 0.10    & 0.38                & 0.19    & 0.73 \\
		H--L    &-0.06    &-0.29                  & 0.31*** & 1.36               & 0.23**  & 1.03                & 0.25*** & 1.22 \\
		$\Delta$& 0.44*** &3.66***                & 0.06    & 2.02***            & 0.15*   & 2.34***             & 0.12    & 2.15***\\
		Turnover& \multicolumn{2}{c}{154\%}       & \multicolumn{2}{c}{391\%}    & \multicolumn{2}{c}{685\%}     & \multicolumn{2}{c}{417\%} \\
		\bottomrule
	\end{tabular*}
\end{table}

\subsection{Factor-adjusted performance}
The return spreads reported above may partly reflect systematic factor exposures rather than abnormal portfolio performance.
To assess this possibility,
we estimate the Fama--French three-factor model (FF3) \citep{FamaFrench1993} and the Fama--French five-factor model (FF5) \citep{FamaFrench2015} for each H--L portfolio.
FF3 includes the market excess-return factor (MKT--RF), the size factor (small minus big, SMB), and the value factor (high minus low, HML).
FF5 augments these factors with the profitability factor (robust minus weak, RMW) and the investment factor (conservative minus aggressive, CMA).
We use daily Chinese A-share factor returns from the China Stock Market \& Accounting Research (CSMAR) database (\url{https://data.csmar.com/}).
The detailed construction of the factors follows the two original studies \citep{FamaFrench1993,FamaFrench2015}.

For each portfolio formation date,
daily factor returns from the first through the fifth subsequent market trading days are compounded to match the five-trading-day holding period of the portfolio returns.
We regress each H--L portfolio return on the corresponding FF3 or FF5 factors,
with alpha defined as the regression intercept and annualized consistently with the portfolio returns.
Because H--L is a zero-cost portfolio, its return is used directly as the dependent variable without subtracting the risk-free rate.
For each alternative specification or traditional strategy $j$,
we apply the same factor regression to the paired H--L return difference, defined as the return of the OHLC-only portfolio minus that of the comparator.
The intercept from this regression is $\Delta\alpha_j=\alpha_{\mathrm{OHLC}}-\alpha_j$, so a positive value indicates higher factor-adjusted performance by the OHLC-only 2D CNN.
Estimating the paired return difference directly accounts for the contemporaneous dependence between the two portfolios and tests whether their alphas differ.
The significance of both the alpha estimates and alpha differences is assessed using two-sided tests with Newey--West standard errors and one lag,
consistent with the return tests described above.

Table~\ref{tab:2DCNN_alpha} reports the annualized alpha estimates and alpha differences.
The FF3 and FF5 estimates are nearly identical throughout the table,
indicating that adding the profitability and investment factors does not materially alter the estimated abnormal performance of these short-horizon portfolios.

Panel A shows that all five individual feature-group specifications generate positive alphas that are significant at the 1\% level under both factor models.
The OHLC-only specification produces an annualized alpha of 0.38, whereas the estimates for Volume, Trend, Momentum, and Volatility range from 0.17 to 0.27.
The corresponding alpha advantages of the OHLC-only model range from 0.11 to 0.22 and are significant at the 5\% level or better in every comparison.
Factor adjustment therefore preserves the conclusion that
the OHLC-only specification yields stronger out-of-sample portfolio performance than any of the alternative feature groups used separately.

Panel B provides the more direct test of whether adding further inputs improves upon OHLC.
The OHLC + Volume and All features specifications generate alphas of 0.41 and 0.39, respectively, slightly above the OHLC-only estimate,
but their alpha differences of $-0.03$ and $-0.01$ are not statistically significant.
Adding Trend or Momentum lowers alpha to 0.33 and 0.32;
the corresponding OHLC-minus-alternative differences of 0.05 and 0.06 are significant at the 10\% level under both factor models.
The OHLC + Volatility specification produces an alpha of 0.36, and its difference from OHLC is also insignificant.
Consequently, none of the augmented-input specifications delivers a statistically significant improvement in factor-adjusted performance over the OHLC-only model.

Panel C compares the OHLC-only model with the traditional strategies.
MOM has alpha estimates of $-0.04$ under FF3 and $-0.06$ under FF5, neither of which is statistically significant,
whereas STR, WSTR, and TREND generate significantly positive alphas.
The OHLC-only model has a higher alpha point estimate than all four strategies.
Its alpha advantages over MOM, WSTR, and TREND are statistically significant, but the 0.09 difference relative to STR is not.
The evidence therefore does not imply uniform alpha dominance over every traditional strategy:
relative to STR, the OHLC-only model has a higher alpha point estimate, but the difference cannot be statistically distinguished from zero.
Nevertheless, as shown in Table~\ref{tab:traditional_performance}, the OHLC-only model achieves a significantly higher Sharpe ratio than STR.

Overall, controlling for common equity factors leaves the principal portfolio findings unchanged.
The OHLC-only specification produces substantially larger alphas than the standalone alternative feature groups and
generally stronger factor-adjusted performance than the traditional strategies.
More importantly for the central question of this study,
adding trading volume or technical-indicator groups to OHLC does not produce a statistically significant increase in alpha.

Taken together,
the return, Sharpe ratio, and factor-adjusted alpha comparisons consistently show that
explicitly augmenting the OHLC input with trading volume or technical-indicator groups does not produce a statistically detectable improvement in out-of-sample portfolio performance.
This conclusion concerns the incremental economic value of the additional inputs rather than the internal representations learned by the 2D CNN.
\begin{table}[H]
	\centering
	\caption{\textbf{Factor-adjusted H--L portfolio performance based on 2D CNN forecasts and traditional trading strategies.}
	This table reports annualized FF3 and FF5 alpha estimates for the H--L portfolios reported in Tables~\ref{tab:2dcnn} and~\ref{tab:traditional_performance}.
	Panels A and B present 2D CNN specifications using individual and combined feature groups, respectively,
	while Panel C presents the MOM, STR, WSTR, and TREND strategies.
	The columns labeled $\Delta\alpha$ report the corresponding alpha differences relative to the OHLC-only 2D CNN specification.
	***, **, and * denote significance at the 1\%, 5\%, and 10\% levels, respectively.}
	\label{tab:2DCNN_alpha}
	\begin{tabular*}{\textwidth}{@{\extracolsep{\fill}} l cccc}
		\toprule
		\multirow{2}{*}{Model specification / traditional strategy}
		& \multicolumn{2}{c}{$\alpha$}
		& \multicolumn{2}{c}{$\Delta\alpha$} \\
		\cmidrule(lr){2-3} \cmidrule(lr){4-5}
		& FF3 & FF5 & FF3 & FF5 \\
		\midrule
		\multicolumn{5}{l}{Panel A: Individual feature groups}\\
		\midrule
		OHLC              & 0.38***     & 0.38***      & --         & --      \\
		Volume            & 0.24***     & 0.24***      & 0.14**     & 0.14**  \\
		Trend             & 0.27***     & 0.27***      & 0.11**     & 0.11**  \\
		Momentum          & 0.25***     & 0.25***      & 0.13***    & 0.13*** \\
		Volatility        & 0.17***     & 0.17***      & 0.22***    & 0.21*** \\

		\midrule
		\multicolumn{5}{l}{Panel B: Combined feature groups}\\
		\midrule
		OHLC + Volume     & 0.41***     & 0.41***      &-0.03       &-0.03     \\
		OHLC + Trend      & 0.33***     & 0.33***      & 0.05*      & 0.05*    \\
		OHLC + Momentum   & 0.32***     & 0.32***      & 0.06*      & 0.06*    \\
		OHLC + Volatility & 0.36***     & 0.36***      & 0.03       & 0.02     \\
		All features      & 0.39***     & 0.39***      &-0.01       &-0.01     \\

		\midrule
		\multicolumn{5}{l}{Panel C: Traditional trading strategies}\\
		\midrule
		MOM               &-0.04        &-0.06         & 0.43***    & 0.44*** \\
		STR               & 0.29***     & 0.29***      & 0.09       & 0.09    \\
		WSTR              & 0.18**      & 0.17**       & 0.21**     & 0.21**  \\
		TREND             & 0.23**      & 0.22**       & 0.16*      & 0.16*   \\
		\bottomrule
	\end{tabular*}
\end{table}

\section{Portfolio performance based on numerical-input models}
\label{sec:portfolio_numerical}
This section examines whether the principal findings from the 2D CNN extend to logistic regression, 1D CNN,
and LSTM models using numerical representations of the market data underlying the images.

Throughout this section, decile portfolio formation, performance measurement, factor adjustment,
and statistical inference follow the procedures described in Section~\ref{sec:portfolio_2DCNN}.
For each model,
the corresponding out-of-sample predicted probabilities are used to sort the same stock sample on the same portfolio formation dates
into equal-weighted deciles with non-overlapping scheduled five-trading-day holding periods.
Ret, SR, Turnover, H--L, $\Delta$, $\alpha$, and $\Delta\alpha$ retain the definitions used in Section~\ref{sec:portfolio_2DCNN}.
The statistical tests and significance conventions are also unchanged.

\subsection{Logistic regression results}
We begin with the $\ell_2$-regularized logistic regression,
which provides a linear benchmark for assessing whether the contribution of the alternative feature groups depends on a nonlinear model architecture.
Table~\ref{tab:logit} reports the resulting out-of-sample decile portfolio performance.

Across both panels of Table~\ref{tab:logit},
High earns a higher return than Low under every input specification,
although the intermediate deciles do not always exhibit a perfectly monotonic pattern.
All H--L portfolios generate positive annualized returns that are significant at the 1\% level,
indicating that each input specification produces economically meaningful separation in realized returns between stocks with relatively high and low predicted probabilities.

Panel A of Table~\ref{tab:logit} presents the individual feature groups.
The OHLC-only specification produces an annualized H--L return of 0.30 and a Sharpe ratio of 1.56.
The corresponding H--L returns for Volume, Trend, Momentum, and Volatility are 0.19, 0.26, 0.25, and 0.28, respectively, with Sharpe ratios ranging from 1.32 to 1.57.
Although the OHLC-only specification has the largest H--L return point estimate,
none of its return or Sharpe-ratio differences relative to the other individual feature groups is statistically significant.
The logistic regression therefore does not establish a significant performance advantage of OHLC over the alternative feature groups used separately.

Panel B of Table~\ref{tab:logit} examines the combined feature groups.
The augmented specifications generate H--L returns between 0.25 and 0.30 and Sharpe ratios between 1.50 and 1.69.
OHLC + Momentum and All features both produce an H--L return of 0.30, equal to that of the OHLC-only specification, while their Sharpe ratios are slightly higher.
Nevertheless, none of the differences in H--L returns or Sharpe ratios is statistically significant.
Explicitly adding Volume, Trend, Momentum, Volatility,
or all additional features therefore produces no statistically detectable improvement in portfolio performance under the linear specification.

Average monthly turnover ranges from 481\% to 647\% among the individual feature groups and from 498\% to 636\% among the combined specifications,
confirming that trading intensity remains high across the logistic regression portfolios.
\begin{table}[H]
	\centering
	\caption{\textbf{Decile portfolio performance based on $\ell_2$-regularized logistic regression forecasts.}
	This table reports the out-of-sample performance of equal-weighted decile portfolios across alternative numerical input specifications.
	Panel A presents individual feature groups,
	whereas Panel B presents OHLC combined with additional inputs,
	including an All features specification containing every input group.
	Ret and SR denote the annualized mean return and annualized Sharpe ratio, respectively,
	while Turnover denotes the average monthly turnover of the H--L portfolio.
	The $\Delta$ row reports the corresponding performance differences relative to the OHLC-only specification.
	***, **, and * denote significance at the 1\%, 5\%, and 10\% levels, respectively.}
	\label{tab:logit}
	\begin{tabular*}{\textwidth}{@{\extracolsep{\fill}} l *{5}{cc}}
		\toprule
		\multicolumn{11}{l}{Panel A: Individual feature groups}\\
		\midrule
		&\multicolumn{2}{c}{OHLC}
		&\multicolumn{2}{c}{Volume}
		&\multicolumn{2}{c}{Trend}
		&\multicolumn{2}{c}{Momentum}
		&\multicolumn{2}{c}{Volatility} \\
		        \cmidrule(lr){2-3}            \cmidrule(lr){4-5}             \cmidrule(lr){6-7}             \cmidrule(lr){8-9}             \cmidrule(lr){10-11}
		        & Ret     & SR                & Ret     & SR                 & Ret     & SR                 & Ret     & SR                 & Ret     & SR \\
		\midrule
		Low     &-0.08    &-0.38              & 0.00    & 0.02               &-0.03    &-0.12               &-0.05    &-0.24               &-0.05    &-0.21\\
		2       & 0.09    & 0.46              & 0.17    & 0.80               & 0.13    & 0.65               & 0.11    & 0.53               & 0.13    & 0.63\\
		3       & 0.15    & 0.75              & 0.16    & 0.76               & 0.13    & 0.65               & 0.16    & 0.78               & 0.13    & 0.63\\
		4       & 0.17    & 0.84              & 0.18    & 0.84               & 0.15    & 0.71               & 0.17    & 0.85               & 0.16    & 0.78\\
		5       & 0.19    & 0.94              & 0.16    & 0.77               & 0.18    & 0.85               & 0.19    & 0.92               & 0.18    & 0.87\\
		6       & 0.22    & 1.06              & 0.18    & 0.88               & 0.17    & 0.84               & 0.22    & 1.05               & 0.19    & 0.92\\
		7       & 0.21    & 0.97              & 0.19    & 0.91               & 0.20    & 0.94               & 0.21    & 1.00               & 0.20    & 0.94\\
		8       & 0.23    & 1.08              & 0.19    & 0.92               & 0.19    & 0.89               & 0.21    & 0.94               & 0.22    & 1.02\\
		9       & 0.22    & 1.00              & 0.20    & 1.02               & 0.26    & 1.19               & 0.20    & 0.87               & 0.23    & 1.01\\
		High    & 0.21    & 0.90              & 0.20    & 0.96               & 0.24    & 1.06               & 0.20    & 0.84               & 0.24    & 1.01\\
		H--L    & 0.30*** & 1.56              & 0.19*** & 1.53               & 0.26*** & 1.57               & 0.25*** & 1.32               & 0.28*** & 1.52\\
		$\Delta$& --      & --                & 0.11    & 0.03               & 0.03    &-0.01               & 0.05    & 0.24               & 0.02    & 0.04\\
		Turnover& \multicolumn{2}{c}{647\%}   & \multicolumn{2}{c}{527\%}    & \multicolumn{2}{c}{481\%}    & \multicolumn{2}{c}{622\%}    & \multicolumn{2}{c}{604\%} \\

		\midrule
		\multicolumn{11}{l}{Panel B: Combined feature groups}\\
		\midrule
		&\multicolumn{2}{c}{OHLC + Volume}
		&\multicolumn{2}{c}{OHLC + Trend}
		&\multicolumn{2}{c}{OHLC + Momentum}
		&\multicolumn{2}{c}{OHLC + Volatility}
		&\multicolumn{2}{c}{All features} \\
		        \cmidrule(lr){2-3}            \cmidrule(lr){4-5}            \cmidrule(lr){6-7}            \cmidrule(lr){8-9}            \cmidrule(lr){10-11}
		        & Ret     & SR                & Ret     & SR                & Ret     & SR                   & Ret     & SR                   & Ret     & SR \\
		\midrule
		Low     &-0.05    &-0.24              &-0.03    &-0.12              &-0.05    &-0.26                 &-0.03    &-0.13                 &-0.05    &-0.22  \\
		2       & 0.11    & 0.54              & 0.10    & 0.47              & 0.08    & 0.42                 & 0.10    & 0.50                 & 0.12    & 0.56  \\
		3       & 0.14    & 0.66              & 0.15    & 0.71              & 0.15    & 0.73                 & 0.13    & 0.66                 & 0.12    & 0.58  \\
		4       & 0.19    & 0.89              & 0.16    & 0.76              & 0.17    & 0.82                 & 0.16    & 0.76                 & 0.16    & 0.78  \\
		5       & 0.20    & 0.97              & 0.17    & 0.80              & 0.18    & 0.88                 & 0.18    & 0.86                 & 0.16    & 0.78  \\
		6       & 0.20    & 0.93              & 0.19    & 0.93              & 0.21    & 1.01                 & 0.19    & 0.92                 & 0.19    & 0.91  \\
		7       & 0.20    & 0.98              & 0.18    & 0.88              & 0.21    & 1.00                 & 0.19    & 0.93                 & 0.21    & 0.97  \\
		8       & 0.23    & 1.07              & 0.21    & 0.97              & 0.20    & 0.93                 & 0.20    & 0.93                 & 0.21    & 0.98  \\
		9       & 0.23    & 1.05              & 0.25    & 1.15              & 0.23    & 0.99                 & 0.24    & 1.07                 & 0.26    & 1.16  \\
		High    & 0.20    & 0.88              & 0.25    & 1.12              & 0.24    & 1.03                 & 0.25    & 1.06                 & 0.25    & 1.11  \\
		H--L    & 0.25*** & 1.57              & 0.28*** & 1.64              & 0.30*** & 1.69                 & 0.28*** & 1.50                 & 0.30*** & 1.64  \\
		$\Delta$& 0.05    &-0.01              & 0.02    &-0.07              & 0.00    &-0.13                 & 0.02    & 0.06                 & 0.00    &-0.08\\
		Turnover& \multicolumn{2}{c}{556\%}   & \multicolumn{2}{c}{521\%}   & \multicolumn{2}{c}{636\%}      & \multicolumn{2}{c}{594\%}      & \multicolumn{2}{c}{498\%} \\
		\bottomrule
	\end{tabular*}
\end{table}

Table~\ref{tab:logit_alpha} reports the corresponding factor-adjusted performance.
The FF3 and FF5 estimates are closely aligned throughout the table,
indicating that adding the profitability and investment factors does not materially alter the estimated abnormal performance of these short-horizon portfolios.

Panel A of Table~\ref{tab:logit_alpha} shows that all individual feature groups produce positive alphas that are significant at the 1\% level under both factor models.
The OHLC-only specification generates annualized alphas of 0.29 under FF3 and 0.28 under FF5, while the estimates for the other individual feature groups range from 0.22 to 0.30.
None of the corresponding alpha differences is statistically significant.
Factor adjustment therefore confirms that the performance of the OHLC-only logistic regression cannot be distinguished statistically from
that of the alternative feature groups used separately.

Panel B of Table~\ref{tab:logit_alpha} yields a similar conclusion for the combined specifications.
Their annualized alphas range from 0.25 to 0.33 and are significant at the 1\% level under both factor models.
The All features specification has the largest alpha point estimate, at 0.33 under both FF3 and FF5, compared with 0.29 and 0.28 for the OHLC-only specification.
However, the corresponding alpha differences of $-0.04$ and $-0.05$ are not statistically significant.
The alpha differences for the other augmented specifications are also insignificant.
Thus, none of the augmented specifications produces a statistically detectable improvement in factor-adjusted performance.

Panel C of Table~\ref{tab:logit_alpha} compares the OHLC-only logistic regression with the traditional strategies.
Its alpha advantages over MOM and WSTR are significant at the 1\% and 5\% levels, respectively, whereas the differences relative to STR and TREND are not statistically significant.
The OHLC-only logistic regression therefore generally exhibits stronger factor-adjusted performance than the traditional strategies.
\begin{table}[H]
	\centering
	\caption{\textbf{Factor-adjusted H--L portfolio performance based on $\ell_2$-regularized logistic regression forecasts and traditional trading strategies.}
	This table reports annualized FF3 and FF5 alpha estimates for the H--L portfolios reported in Tables~\ref{tab:logit} and~\ref{tab:traditional_performance}.
	Panels A and B present $\ell_2$-regularized logistic regression specifications using individual and combined feature groups, respectively,
	while Panel C presents the MOM, STR, WSTR, and TREND strategies.
	The columns labeled $\Delta\alpha$ report the corresponding alpha differences relative to the OHLC-only $\ell_2$-regularized logistic regression specification.
	***, **, and * denote significance at the 1\%, 5\%, and 10\% levels, respectively.}
	\label{tab:logit_alpha}
	\begin{tabular*}{\textwidth}{@{\extracolsep{\fill}} l cccc}
		\toprule
		\multirow{2}{*}{Model specification / traditional strategy}
		& \multicolumn{2}{c}{$\alpha$}
		& \multicolumn{2}{c}{$\Delta\alpha$} \\
		\cmidrule(lr){2-3} \cmidrule(lr){4-5}
		& FF3 & FF5 & FF3 & FF5 \\
		\midrule
		\multicolumn{5}{l}{Panel A: Individual feature groups}\\
		\midrule
		OHLC              & 0.29***     & 0.28***      & --         & --  \\
		Volume            & 0.24***     & 0.22***      & 0.05    & 0.06   \\
		Trend             & 0.29***     & 0.29***      & 0.00    &-0.01   \\
		Momentum          & 0.23***     & 0.22***      & 0.06    & 0.06   \\
		Volatility        & 0.29***     & 0.30***      & 0.00    &-0.02   \\

		\midrule
		\multicolumn{5}{l}{Panel B: Combined feature groups}\\
		\midrule
		OHLC + Volume     & 0.26***     & 0.25***     & 0.03     & 0.03   \\
		OHLC + Trend      & 0.30***     & 0.30***     &-0.01     &-0.02   \\
		OHLC + Momentum   & 0.28***     & 0.28***     & 0.01     & 0.00   \\
		OHLC + Volatility & 0.28***     & 0.28***     & 0.01     & 0.00   \\
		All features      & 0.33***     & 0.33***     &-0.04     &-0.05   \\

		\midrule
		\multicolumn{5}{l}{Panel C: Traditional trading strategies}\\
		\midrule
		MOM               &-0.04        &-0.06        & 0.33***  & 0.34***\\
		STR               & 0.29***     & 0.29***     & 0.00     &-0.01   \\
		WSTR              & 0.18**      & 0.17**      & 0.11**   & 0.11** \\
		TREND             & 0.23**      & 0.22**      & 0.06     & 0.06   \\
		\bottomrule
	\end{tabular*}
\end{table}

Taken together, the logistic regression results provide two findings.
First, all input specifications generate significantly positive return spreads, but OHLC does not significantly outperform the standalone alternative feature groups.
Second, explicitly augmenting OHLC with trading volume or technical indicators does not significantly improve H--L returns, Sharpe ratios, or factor-adjusted alphas.
The absence of a statistically significant incremental portfolio-performance improvement is therefore not unique to the nonlinear 2D CNN or its image-based representation.

\subsection{1D CNN results}
We next examine the 1D CNN, which applies convolution directly to the temporal dimension of the numerical inputs.
Table~\ref{tab:1dcnn} reports its out-of-sample decile portfolio performance.

Across both panels of Table~\ref{tab:1dcnn},
High earns a higher return than Low under every input specification,
although the intermediate deciles are not perfectly monotonic in all cases.
All H--L portfolios generate positive annualized returns that are significant at the 1\% level,
indicating that the 1D CNN produces economically meaningful separation in realized returns between stocks with relatively high and low predicted probabilities.

Panel A of Table~\ref{tab:1dcnn} presents the individual feature groups.
The OHLC-only specification generates an annualized H--L return of 0.34 and a Sharpe ratio of 3.09.
By comparison, Volume, Trend, Momentum, and Volatility produce H--L returns of 0.19, 0.24, 0.19, and 0.19, respectively, with Sharpe ratios ranging from 1.39 to 2.16.
The OHLC-only specification has significantly higher H--L returns and Sharpe ratios than all four alternative feature groups.
Its return advantages range from 0.09 to 0.15, while its Sharpe-ratio advantages range from 0.92 to 1.69.
All eight differences are significant at the 10\% level or better.
Thus, when the feature groups are used separately, the 1D CNN extracts substantially greater economic value from OHLC than from Volume, Trend, Momentum, or Volatility.

Panel B of Table~\ref{tab:1dcnn} examines the combined feature groups.
The augmented specifications generate annualized H--L returns ranging from 0.30 to 0.39 and Sharpe ratios ranging from 2.56 to 3.62.
Adding Volume raises the H--L return to 0.36, while adding Trend, Momentum, or Volatility produces returns of 0.32, 0.31, and 0.30, respectively.
The All features specification has the largest point estimates, with an H--L return of 0.39 and a Sharpe ratio of 3.62, compared with 0.34 and 3.09 for the OHLC-only specification.
However, none of these augmented specifications delivers a statistically significant increase in either H--L return or Sharpe ratio.
Moreover, OHLC + Volatility has a significantly lower Sharpe ratio than the OHLC-only specification at the 10\% level.
The combined-input results therefore provide no statistically significant evidence that explicitly adding the alternative feature groups improves portfolio performance.
\begin{table}[H]
	\centering
	\caption{\textbf{Decile portfolio performance based on 1D CNN forecasts.}
	This table reports the out-of-sample performance of equal-weighted decile portfolios across alternative input specifications.
	Panel A presents individual feature groups,
	whereas Panel B presents OHLC combined with additional inputs,
	including an All features specification containing every input group.
	Ret and SR denote the annualized mean return and annualized Sharpe ratio, respectively,
	while Turnover denotes the average monthly turnover of the H--L portfolio.
	The $\Delta$ row reports the corresponding performance differences relative to the OHLC-only specification.
	***, **, and * denote significance at the 1\%, 5\%, and 10\% levels, respectively.}
	\label{tab:1dcnn}
	\begin{tabular*}{\textwidth}{@{\extracolsep{\fill}} l *{5}{cc}}
		\toprule
		\multicolumn{11}{l}{Panel A: Individual feature groups}\\
		\midrule
		&\multicolumn{2}{c}{OHLC}
		&\multicolumn{2}{c}{Volume}
		&\multicolumn{2}{c}{Trend}
		&\multicolumn{2}{c}{Momentum}
		&\multicolumn{2}{c}{Volatility} \\
		         \cmidrule(lr){2-3}            \cmidrule(lr){4-5}             \cmidrule(lr){6-7}             \cmidrule(lr){8-9}             \cmidrule(lr){10-11}
		         & Ret     & SR                & Ret     & SR                 & Ret     & SR                 & Ret     & SR                  & Ret     & SR \\
		\midrule
		Low      &-0.05    &-0.22              &-0.01    &-0.04               & 0.03    & 0.13               & 0.06    & 0.32                & 0.06    & 0.27   \\
		2        & 0.05    & 0.25              & 0.15    & 0.70               & 0.07    & 0.35               & 0.09    & 0.43                & 0.08    & 0.39   \\
		3        & 0.12    & 0.59              & 0.17    & 0.81               & 0.09    & 0.45               & 0.09    & 0.43                & 0.13    & 0.64   \\
		4        & 0.15    & 0.71              & 0.18    & 0.87               & 0.15    & 0.72               & 0.12    & 0.58                & 0.13    & 0.62   \\
		5        & 0.17    & 0.84              & 0.19    & 0.92               & 0.16    & 0.78               & 0.14    & 0.66                & 0.15    & 0.73   \\
		6        & 0.21    & 1.00              & 0.18    & 0.88               & 0.17    & 0.84               & 0.18    & 0.85                & 0.19    & 0.94   \\
		7        & 0.20    & 0.99              & 0.21    & 1.04               & 0.20    & 0.98               & 0.22    & 1.05                & 0.18    & 0.86   \\
		8        & 0.23    & 1.10              & 0.18    & 0.92               & 0.23    & 1.12               & 0.21    & 1.02                & 0.22    & 1.06   \\
		9        & 0.25    & 1.20              & 0.19    & 0.96               & 0.24    & 1.15               & 0.26    & 1.21                & 0.24    & 1.08   \\
		High     & 0.29    & 1.34              & 0.18    & 0.89               & 0.27    & 1.23               & 0.26    & 1.17                & 0.25    & 1.04   \\
		H--L     & 0.34*** & 3.09              & 0.19*** & 1.57               & 0.24*** & 2.16               & 0.19*** & 1.83                & 0.19*** & 1.39   \\
		$\Delta$ & --      & --                & 0.15**  & 1.52**             & 0.09*   & 0.92*              & 0.14*** & 1.25***             & 0.15*** & 1.69***\\
		Turnover & \multicolumn{2}{c}{667\%}   & \multicolumn{2}{c}{624\%}    & \multicolumn{2}{c}{625\%}    & \multicolumn{2}{c}{663\%}     & \multicolumn{2}{c}{664\%} \\

		\midrule
		\multicolumn{11}{l}{Panel B: Combined feature groups}\\
		\midrule
		&\multicolumn{2}{c}{OHLC + Volume}
		&\multicolumn{2}{c}{OHLC + Trend}
		&\multicolumn{2}{c}{OHLC + Momentum}
		&\multicolumn{2}{c}{OHLC + Volatility}
		&\multicolumn{2}{c}{All features} \\
		        \cmidrule(lr){2-3}            \cmidrule(lr){4-5}            \cmidrule(lr){6-7}            \cmidrule(lr){8-9}            \cmidrule(lr){10-11}
		        & Ret     & SR                & Ret     & SR                & Ret     & SR                & Ret     & SR                 & Ret     & SR \\
		\midrule
		Low     &-0.09    &-0.40              &-0.04    &-0.18              & 0.00    & 0.00              & 0.01    & 0.07               &-0.09    &-0.41  \\
		2       & 0.08    & 0.35              & 0.04    & 0.21              & 0.04    & 0.18              & 0.06    & 0.28               & 0.06    & 0.30  \\
		3       & 0.12    & 0.55              & 0.10    & 0.49              & 0.10    & 0.49              & 0.09    & 0.44               & 0.10    & 0.47  \\
		4       & 0.16    & 0.80              & 0.13    & 0.63              & 0.12    & 0.58              & 0.10    & 0.48               & 0.12    & 0.57  \\
		5       & 0.19    & 0.92              & 0.16    & 0.80              & 0.19    & 0.91              & 0.16    & 0.78               & 0.17    & 0.84  \\
		6       & 0.21    & 1.03              & 0.18    & 0.87              & 0.17    & 0.84              & 0.19    & 0.91               & 0.21    & 1.04  \\
		7       & 0.22    & 1.09              & 0.22    & 1.12              & 0.23    & 1.11              & 0.22    & 1.10               & 0.24    & 1.20  \\
		8       & 0.23    & 1.10              & 0.26    & 1.28              & 0.23    & 1.13              & 0.23    & 1.11               & 0.23    & 1.10  \\
		9       & 0.25    & 1.18              & 0.27    & 1.29              & 0.24    & 1.14              & 0.25    & 1.19               & 0.28    & 1.31  \\
		High    & 0.27    & 1.25              & 0.29    & 1.32              & 0.31    & 1.40              & 0.31    & 1.39               & 0.30    & 1.35  \\
		H--L    & 0.36*** & 3.20              & 0.32*** & 3.24              & 0.31*** & 2.97              & 0.30*** & 2.56               & 0.39*** & 3.62  \\
		$\Delta$&-0.03    &-0.11              & 0.01    &-0.15              & 0.03    & 0.11              & 0.04    & 0.53*              & -0.05   &-0.54  \\
		Turnover& \multicolumn{2}{c}{666\%}   & \multicolumn{2}{c}{655\%}   & \multicolumn{2}{c}{669\%}   & \multicolumn{2}{c}{667\%}    & \multicolumn{2}{c}{653\%} \\
		\bottomrule
	\end{tabular*}
\end{table}

Average monthly turnover ranges from 624\% to 667\% among the individual feature groups and from 653\% to 669\% among the combined specifications.
Trading intensity is therefore uniformly high and varies relatively little across the 1D CNN specifications.

Table~\ref{tab:1DCNN_alpha} reports the corresponding factor-adjusted performance.
The FF3 and FF5 alpha estimates are again closely aligned.

Panel A of Table~\ref{tab:1DCNN_alpha} shows that all individual feature groups generate positive alphas that are significant at the 1\% level under both factor models.
The OHLC-only specification produces annualized alphas of 0.35 under FF3 and 0.34 under FF5.
The corresponding estimates for Volume, Trend, Momentum, and Volatility range from 0.18 to 0.25.
The alpha advantages of the OHLC-only specification range from 0.10 to 0.17 and are significant at the 10\% level or better in all four comparisons.
Factor adjustment therefore preserves the conclusion that OHLC produces stronger standalone portfolio performance than the alternative feature groups.

Panel B of Table~\ref{tab:1DCNN_alpha} presents the combined specifications.
Their annualized alphas range from 0.29 to 0.40 and are significant at the 1\% level under both factor models.
The All features specification again has the largest point estimates, at 0.40 under FF3 and 0.39 under FF5, compared with 0.35 and 0.34 for the OHLC-only specification.
Nevertheless, its alpha differences of $-0.05$ under both factor models are not statistically significant.
The alpha differences for the other augmented specifications are also insignificant.
Consequently, none of the additional input groups produces a statistically detectable improvement in factor-adjusted performance.

Panel C of Table~\ref{tab:1DCNN_alpha} compares the OHLC-only 1D CNN with the traditional strategies.
Its alpha advantages over MOM are significant at the 1\% level, while those relative to WSTR are significant at the 10\% level.
The differences relative to STR and TREND are not statistically significant.
The OHLC-only 1D CNN therefore generally exhibits stronger factor-adjusted performance than the traditional strategies.

Taken together, the 1D CNN results reinforce the principal findings obtained from the 2D CNN.
OHLC produces significantly stronger standalone portfolio performance than Volume, Trend, Momentum, or Volatility,
whereas explicitly adding these inputs to OHLC does not generate a statistically detectable improvement in H--L returns, Sharpe ratios, or factor-adjusted alphas.
This conclusion therefore remains valid when the image representation is replaced by numerical sequences processed using one-dimensional convolution.
\begin{table}[H]
	\centering
	\caption{\textbf{Factor-adjusted H--L portfolio performance based on 1D CNN forecasts and traditional trading strategies.}
	This table reports annualized FF3 and FF5 alpha estimates for the H--L portfolios reported in Tables~\ref{tab:1dcnn} and~\ref{tab:traditional_performance}.
	Panels A and B present 1D CNN specifications using individual and combined feature groups, respectively,
	while Panel C presents the MOM, STR, WSTR, and TREND strategies.
	The columns labeled $\Delta\alpha$ report the corresponding alpha differences relative to the OHLC-only 1D CNN specification.
	***, **, and * denote significance at the 1\%, 5\%, and 10\% levels, respectively.}
	\label{tab:1DCNN_alpha}
	\begin{tabular*}{\textwidth}{@{\extracolsep{\fill}} l cccc}
		\toprule
		\multirow{2}{*}{Model specification / traditional strategy}
		& \multicolumn{2}{c}{$\alpha$}
		& \multicolumn{2}{c}{$\Delta\alpha$} \\
		\cmidrule(lr){2-3} \cmidrule(lr){4-5}
		& FF3 & FF5 & FF3 & FF5 \\
		\midrule
		\multicolumn{5}{l}{Panel A: Individual feature groups}\\
		\midrule
		OHLC              & 0.35***    & 0.34***    & --        & --     \\
		Volume            & 0.24***    & 0.24***    & 0.11*     & 0.11*  \\
		Trend             & 0.25***    & 0.24***    & 0.10*     & 0.10*  \\
		Momentum          & 0.18***    & 0.18***    & 0.17***   & 0.17***\\
		Volatility        & 0.19***    & 0.19***    & 0.16***   & 0.16***\\

		\midrule
		\multicolumn{5}{l}{Panel B: Combined feature groups}\\
		\midrule
		OHLC + Volume     & 0.37***    & 0.37***    &-0.02      &-0.02   \\
		OHLC + Trend      & 0.33***    & 0.32***    & 0.02      & 0.02   \\
		OHLC + Momentum   & 0.30***    & 0.29***    & 0.05      & 0.05   \\
		OHLC + Volatility & 0.30***    & 0.30***    & 0.05      & 0.05   \\
		All features      & 0.40***    & 0.39***    &-0.05      &-0.05   \\

		\midrule
		\multicolumn{5}{l}{Panel C: Traditional trading strategies}\\
		\midrule
		MOM               &-0.04        &-0.06      & 0.39***   & 0.41***\\
		STR               & 0.29***     & 0.29***   & 0.06      & 0.06   \\
		WSTR              & 0.18**      & 0.17**    & 0.17*     & 0.17*  \\
		TREND             & 0.23**      & 0.22**    & 0.12      & 0.13   \\
		\bottomrule
	\end{tabular*}
\end{table}

\subsection{LSTM results}
We next examine the LSTM, which processes the numerical inputs as temporal sequences.
Table~\ref{tab:lstm} reports its out-of-sample decile portfolio performance.

Across both panels of Table~\ref{tab:lstm},
High earns a higher return than Low under every input specification, although the intermediate deciles are not perfectly monotonic in all cases.
All H--L portfolios generate positive annualized returns that are significant at the 5\% level or better,
indicating that the LSTM forecasts provide meaningful separation in realized returns between stocks with relatively high and low predicted probabilities.

Panel A of Table~\ref{tab:lstm} presents the individual feature groups.
The OHLC-only specification produces an annualized H--L return of 0.34 and a Sharpe ratio of 2.10.
The corresponding H--L returns for Volume, Trend, Momentum, and Volatility are 0.27, 0.26, 0.26, and 0.19, respectively, with Sharpe ratios ranging from 1.15 to 2.01.
Although the OHLC-only specification has the largest point estimates for both H--L return and Sharpe ratio,
the statistical significance of its performance advantage varies across the alternative feature groups.
Its return advantage is significant at the 10\% level relative to Trend and at the 1\% level relative to Volatility, but is insignificant relative to Volume and Momentum.
Its Sharpe-ratio advantage is significant at the 5\% level relative to Momentum and at the 1\% level relative to Volatility, but is insignificant relative to Volume and Trend.
The OHLC-only LSTM therefore generally produces stronger standalone performance,
although this advantage is not statistically significant in every comparison.

Panel B of Table~\ref{tab:lstm} presents the combined feature groups.
The augmented specifications generate annualized H--L returns ranging from 0.24 to 0.35 and Sharpe ratios ranging from 1.75 to 2.86.
Adding Trend or Momentum reduces the H--L return from 0.34 to 0.24 and 0.27, respectively.
The corresponding return differences are significant at the 1\% and 5\% levels.
Adding Volatility produces an H--L return of 0.29 and a Sharpe ratio of 2.09, neither of which differs significantly from the OHLC-only result.
\begin{table}[H]
	\centering
	\caption{\textbf{Decile portfolio performance based on LSTM forecasts.}
	This table reports the out-of-sample performance of equal-weighted decile portfolios across alternative input specifications.
	Panel A presents individual feature groups,
	whereas Panel B presents OHLC combined with additional inputs,
	including an All features specification containing every input group.
	Ret and SR denote the annualized mean return and annualized Sharpe ratio, respectively,
	while Turnover denotes the average monthly turnover of the H--L portfolio.
	The $\Delta$ row reports the corresponding performance differences relative to the OHLC-only specification.
	***, **, and * denote significance at the 1\%, 5\%, and 10\% levels, respectively.}
	\label{tab:lstm}
	\begin{tabular*}{\textwidth}{@{\extracolsep{\fill}} l *{5}{cc}}
		\toprule
		\multicolumn{11}{l}{Panel A: Individual feature groups}\\
		\midrule
		&\multicolumn{2}{c}{OHLC}
		&\multicolumn{2}{c}{Volume}
		&\multicolumn{2}{c}{Trend}
		&\multicolumn{2}{c}{Momentum}
		&\multicolumn{2}{c}{Volatility} \\
		         \cmidrule(lr){2-3}            \cmidrule(lr){4-5}             \cmidrule(lr){6-7}             \cmidrule(lr){8-9}             \cmidrule(lr){10-11}
		         & Ret     & SR                & Ret     & SR                 & Ret     & SR                 & Ret     & SR                  & Ret     & SR \\
		\midrule
		Low      &-0.04    &-0.21              &-0.07    &-0.28              & 0.02    & 0.08              &-0.03    &-0.12               & 0.02    & 0.08   \\
		2        & 0.06    & 0.29              & 0.13    & 0.64              & 0.08    & 0.39              & 0.09    & 0.44               & 0.08    & 0.39   \\
		3        & 0.09    & 0.45              & 0.18    & 0.87              & 0.10    & 0.49              & 0.13    & 0.62               & 0.14    & 0.70   \\
		4        & 0.12    & 0.57              & 0.18    & 0.87              & 0.10    & 0.50              & 0.18    & 0.90               & 0.15    & 0.73   \\
		5        & 0.18    & 0.89              & 0.19    & 0.94              & 0.17    & 0.84              & 0.17    & 0.80               & 0.16    & 0.78   \\
		6        & 0.20    & 0.98              & 0.18    & 0.88              & 0.18    & 0.89              & 0.20    & 0.96               & 0.19    & 0.94   \\
		7        & 0.22    & 1.04              & 0.20    & 1.00              & 0.21    & 0.99              & 0.20    & 0.94               & 0.23    & 1.11   \\
		8        & 0.24    & 1.15              & 0.22    & 1.05              & 0.22    & 1.04              & 0.21    & 0.97               & 0.23    & 1.04   \\
		9        & 0.27    & 1.23              & 0.21    & 1.01              & 0.27    & 1.28              & 0.24    & 1.03               & 0.22    & 0.99   \\
		High     & 0.29    & 1.17              & 0.20    & 0.90              & 0.27    & 1.17              & 0.23    & 0.96               & 0.20    & 0.84   \\
		H--L     & 0.34*** & 2.10              & 0.27*** & 1.78              & 0.26*** & 2.01              & 0.26*** & 1.46               & 0.19**  & 1.15   \\
		$\Delta$ & --      & --                & 0.07    & 0.31              & 0.08*   & 0.09              & 0.08    & 0.64**             & 0.15*** & 0.95***\\
		Turnover & \multicolumn{2}{c}{675\%}   & \multicolumn{2}{c}{620\%}    & \multicolumn{2}{c}{625\%}    & \multicolumn{2}{c}{645\%}     & \multicolumn{2}{c}{664\%} \\

		\midrule
		\multicolumn{11}{l}{Panel B: Combined feature groups}\\
		\midrule
		&\multicolumn{2}{c}{OHLC + Volume}
		&\multicolumn{2}{c}{OHLC + Trend}
		&\multicolumn{2}{c}{OHLC + Momentum}
		&\multicolumn{2}{c}{OHLC + Volatility}
		&\multicolumn{2}{c}{All features} \\
		        \cmidrule(lr){2-3}            \cmidrule(lr){4-5}            \cmidrule(lr){6-7}            \cmidrule(lr){8-9}            \cmidrule(lr){10-11}
		        & Ret     & SR                & Ret     & SR                & Ret     & SR                & Ret     & SR                 & Ret     & SR \\
		\midrule
		Low     &-0.07    &-0.35              & 0.02    & 0.12              &-0.02    &-0.10              &-0.02    &-0.11               &-0.05    &-0.24  \\
		2       & 0.06    & 0.30              & 0.05    & 0.25              & 0.05    & 0.24              & 0.03    & 0.16               & 0.04    & 0.19  \\
		3       & 0.11    & 0.55              & 0.10    & 0.49              & 0.14    & 0.68              & 0.12    & 0.61               & 0.11    & 0.53  \\
		4       & 0.14    & 0.71              & 0.15    & 0.73              & 0.14    & 0.66              & 0.13    & 0.65               & 0.13    & 0.64  \\
		5       & 0.20    & 1.02              & 0.17    & 0.82              & 0.19    & 0.90              & 0.18    & 0.86               & 0.17    & 0.81  \\
		6       & 0.20    & 0.98              & 0.17    & 0.82              & 0.18    & 0.85              & 0.21    & 0.98               & 0.20    & 0.97  \\
		7       & 0.23    & 1.12              & 0.21    & 0.98              & 0.23    & 1.08              & 0.22    & 1.05               & 0.21    & 1.04  \\
		8       & 0.24    & 1.14              & 0.25    & 1.22              & 0.21    & 1.01              & 0.24    & 1.09               & 0.26    & 1.24  \\
		9       & 0.24    & 1.11              & 0.24    & 1.10              & 0.27    & 1.23              & 0.25    & 1.15               & 0.26    & 1.18  \\
		High    & 0.27    & 1.15              & 0.26    & 1.12              & 0.24    & 1.03              & 0.27    & 1.16               & 0.30    & 1.36  \\
		H--L    & 0.34*** & 2.71              & 0.24*** & 1.77              & 0.27*** & 1.75              & 0.29*** & 2.09               & 0.35*** & 2.86  \\
		$\Delta$& 0.00    &-0.61**            & 0.10*** & 0.33              & 0.07**  & 0.34              & 0.04    & 0.00               &-0.01    &-0.77**\\
		Turnover& \multicolumn{2}{c}{654\%}   & \multicolumn{2}{c}{651\%}   & \multicolumn{2}{c}{652\%}   & \multicolumn{2}{c}{664\%}    & \multicolumn{2}{c}{642\%} \\
		\bottomrule
	\end{tabular*}
\end{table}

The results for Volume and All features differ somewhat from those for the other augmented specifications.
OHLC + Volume generates the same H--L return as OHLC alone but raises the Sharpe ratio from 2.10 to 2.71, with the Sharpe-ratio improvement significant at the 5\% level.
The All features specification produces an H--L return of 0.35, which is not significantly different from the OHLC-only return, but increases the Sharpe ratio to 2.86.
This Sharpe-ratio improvement is also significant at the 5\% level.

Because both specifications contain Volume and significantly outperform OHLC alone in terms of the Sharpe ratio, we additionally compare OHLC + Volume directly with All features.
Additional pairwise tests show that the All features specification exceeds OHLC + Volume by only 0.01 in H--L return and by 0.15 in the Sharpe ratio;
neither difference is statistically significant even at the 10\% level.
Thus, adding the Trend, Momentum, and Volatility groups to OHLC + Volume produces no statistically detectable further improvement in either performance measure.
This pattern is consistent with the inclusion of Volume accounting for the Sharpe-ratio improvement observed for the All features specification,
while the technical-indicator groups provide no detectable incremental benefit once Volume is included.

Average monthly turnover ranges from 620\% to 675\% among the individual feature groups and from 642\% to 664\% among the combined specifications.
As with the other model architectures, trading intensity remains uniformly high across the LSTM portfolios.

Table~\ref{tab:lstm_alpha} reports the corresponding factor-adjusted performance.
The FF3 and FF5 alpha estimates are nearly identical throughout the table.
\begin{table}[H]
	\centering
	\caption{\textbf{Factor-adjusted H--L portfolio performance based on LSTM forecasts and traditional trading strategies.}
	This table reports annualized FF3 and FF5 alpha estimates for the H--L portfolios reported in Tables~\ref{tab:lstm} and~\ref{tab:traditional_performance}.
	Panels A and B present LSTM specifications using individual and combined feature groups, respectively,
	while Panel C presents the MOM, STR, WSTR, and TREND strategies.
	The columns labeled $\Delta\alpha$ report the corresponding alpha differences relative to the OHLC-only LSTM specification.
	***, **, and * denote significance at the 1\%, 5\%, and 10\% levels, respectively.}
	\label{tab:lstm_alpha}
	\begin{tabular*}{\textwidth}{@{\extracolsep{\fill}} l cccc}
		\toprule
		\multirow{2}{*}{Model specification / traditional strategy}
		& \multicolumn{2}{c}{$\alpha$}
		& \multicolumn{2}{c}{$\Delta\alpha$} \\
		\cmidrule(lr){2-3} \cmidrule(lr){4-5}
		& FF3 & FF5 & FF3 & FF5 \\
		\midrule
		\multicolumn{5}{l}{Panel A: Individual feature groups}\\
		\midrule
		OHLC              & 0.33***     & 0.33***      & --      & --     \\
		Volume            & 0.30***     & 0.29***      & 0.04    & 0.04   \\
		Trend             & 0.25***     & 0.25***      & 0.08*   & 0.08*  \\
		Momentum          & 0.24***     & 0.24***      & 0.09    & 0.09   \\
		Volatility        & 0.18***     & 0.18***      & 0.15*** & 0.15***\\

		\midrule
		\multicolumn{5}{l}{Panel B: Combined feature groups}\\
		\midrule
		OHLC + Volume     & 0.36***     & 0.36***      &-0.03    &-0.03   \\
		OHLC + Trend      & 0.23***     & 0.23***      & 0.10*** & 0.10***\\
		OHLC + Momentum   & 0.26***     & 0.26***      & 0.07**  & 0.07** \\
		OHLC + Volatility & 0.30***     & 0.30***      & 0.03    & 0.03   \\
		All features      & 0.37***     & 0.37***      &-0.04    &-0.04   \\

		\midrule
		\multicolumn{5}{l}{Panel C: Traditional trading strategies}\\
		\midrule
		MOM               &-0.04        &-0.06         & 0.38*** & 0.39***\\
		STR               & 0.29***     & 0.29***      & 0.04    & 0.04   \\
		WSTR              & 0.18**      & 0.17**       & 0.16    & 0.16   \\
		TREND             & 0.23**      & 0.22**       & 0.11    & 0.11   \\
		\bottomrule
	\end{tabular*}
\end{table}

Panel A of Table~\ref{tab:lstm_alpha} shows that all individual feature groups generate positive alphas that are significant at the 1\% level under both factor models.
The OHLC-only specification produces an annualized alpha of 0.33 under both FF3 and FF5.
The corresponding estimates for Volume, Trend, Momentum, and Volatility range from 0.18 to 0.30.
The alpha advantage of the OHLC-only specification is significant at the 10\% level relative to Trend and at the 1\% level relative to Volatility,
but is insignificant relative to Volume and Momentum.
Factor adjustment therefore continues to favor the OHLC-only specification in point estimates,
although its advantage is statistically significant in only two of the four comparisons.

Panel B of Table~\ref{tab:lstm_alpha} shows that all combined specifications also generate positive alphas that are significant at the 1\% level.
OHLC + Volume and All features produce alphas of 0.36 and 0.37, respectively, compared with 0.33 for the OHLC-only specification.
However, their alpha differences of $-0.03$ and $-0.04$ are not statistically significant.
The improvements in Sharpe ratios associated with these two specifications therefore do not translate into statistically significant improvements in factor-adjusted alpha.

By contrast,
adding Trend or Momentum lowers the annualized alpha to 0.23 and 0.26, respectively.
The corresponding alpha differences of 0.10 and 0.07 are significant at the 1\% and 5\% levels under both factor models.
The OHLC + Volatility specification produces an alpha of 0.30, which is not significantly different from the OHLC-only estimate.
Consequently,
none of the augmented specifications generates a statistically significant increase in factor-adjusted performance,
while the Trend and Momentum additions significantly reduce alpha.

Panel C of Table~\ref{tab:lstm_alpha} compares the OHLC-only LSTM with the traditional strategies.
Its alpha advantage over MOM is significant at the 1\% level under both factor models.
The differences relative to STR, WSTR, and TREND are positive but statistically insignificant.
The OHLC-only LSTM therefore generally exhibits stronger factor-adjusted performance than the traditional strategies.

Taken together, the LSTM results provide qualified support for the findings obtained from the other model architectures.
The OHLC-only specification generally produces stronger standalone performance,
and none of the additional input groups significantly improves the mean H--L return or factor-adjusted alpha.
Adding Trend or Momentum instead significantly reduces both returns and alphas.
Although OHLC + Volume and All features produce significantly higher Sharpe ratios than OHLC alone,
these two augmented specifications do not differ significantly from each other in either H--L return or Sharpe ratio.
The evidence is therefore consistent with the Sharpe-ratio improvement being primarily associated with Volume rather than with the technical-indicator groups.
More broadly, the additional inputs do not deliver a uniform improvement across performance measures or model specifications.

\subsection{Comparison across model architectures}
Combining the 2D CNN evidence in Section~\ref{sec:portfolio_2DCNN} with the numerical-input results above reveals both commonalities and differences across model architectures.
The OHLC-only specifications under all four architectures generate significantly positive H--L returns and factor-adjusted alphas.
Their annualized H--L returns range from 0.30 to 0.38, while their Sharpe ratios range from 1.56 to 3.37.
The 2D CNN produces the largest point estimates of the H--L return, Sharpe ratio, and alpha, with the 1D CNN producing the next-highest Sharpe ratio.
Nevertheless, the significant return spreads generated by the logistic regression,
1D CNN, and LSTM indicate that the economic value of OHLC is not specific to the image representation employed by the 2D CNN.

The relative advantage of OHLC over the alternative feature groups used separately varies across architectures.
Under both the 2D CNN and 1D CNN, OHLC significantly outperforms Volume, Trend, Momentum, and Volatility in H--L returns, Sharpe ratios, and factor-adjusted alphas.
The OHLC-only LSTM also produces the largest point estimates, but its advantages are significant for only some of the alternative feature groups.
Under the logistic regression, none of the corresponding differences is statistically significant.
The absence of a significant OHLC advantage under the logistic regression may partly reflect the limited representational flexibility of its linear specification.

OHLC histories may contain nonlinear interactions and local temporal patterns that a linear model cannot readily capture,
whereas technical indicators embed prespecified transformations of those histories before estimation.
The clearer OHLC advantage under the convolutional models is consistent with these architectures extracting useful patterns directly from OHLC data,
although the results do not conclusively establish this mechanism.

The principal result concerning the incremental value of additional inputs is more consistent.
Across all four architectures,
no augmented specification produces a statistically significant increase in either the mean H--L return or factor-adjusted alpha relative to OHLC alone.
Some additions instead reduce performance.
The technical-indicator groups therefore provide no robust incremental improvement once OHLC is included.

The only statistically significant improvement among the augmented specifications arises in the Sharpe ratios of the LSTM portfolios.
Both OHLC + Volume and All features produce higher Sharpe ratios than OHLC alone, with both improvements significant at the 5\% level.
However, neither specification significantly improves the mean H--L return or factor-adjusted alpha.
Moreover, OHLC + Volume and All features do not differ significantly from each other in either H--L return or Sharpe ratio.
This pattern is consistent with the LSTM Sharpe-ratio improvement being primarily associated with Volume,
while the technical-indicator groups provide no detectable additional benefit once Volume is included.

Overall, across the four models examined in this paper,
model architecture affects the strength of the standalone advantage of OHLC but leaves the central conclusion unchanged:
explicitly adding the technical-indicator groups to OHLC provides no consistent improvement in portfolio returns,
Sharpe ratios, or factor-adjusted alphas under the settings considered.
These results support a conclusion of limited incremental economic value for these indicator groups, but do not establish that technical indicators are generally redundant.

\section{Practical implementation and market frictions}
\label{sec:implementation}
\subsection{Transaction-cost-adjusted performance}
The high turnover generated by the five-trading-day rebalancing strategy makes transaction costs an important consideration in interpreting the gross portfolio results.
We therefore conduct a transaction-cost sensitivity analysis for the baseline OHLC-only 2D CNN strategy.
Consistent with the per-trade cost adjustments used in related image-based return-prediction studies \citep{JiangKellyXiu2023,JeongByunKim2026},
let $c$ denote a proportional one-way transaction cost.
At each rebalancing date, we treat the portfolio formed for that five-trading-day holding period as fully liquidated and re-established.
The High-only portfolio is therefore charged $2c$ per holding period for entry and exit,
whereas the H--L portfolio is charged $4c$ for round-trip trading in both its long and short legs.
Because this full-reconstitution convention does not net positions retained across consecutive rebalancing dates,
it produces a conservative cost adjustment relative to netting retained positions.

Brokerage commissions are account-specific, whereas stamp duty and transfer fees are set uniformly.
Under the fee schedule prevailing at the end of the sample,
stamp duty is levied only on sales at 5 basis points,
while the transfer fee is approximately 0.1 basis points on each side \citep{CSDC2026FeeSchedule}.
Under the symmetric one-way convention adopted here,
these non-commission charges correspond to an equivalent one-way cost of approximately 2.6 basis points.
The 5-basis-point assumption therefore allows approximately 2.4 additional basis points for brokerage commissions and other proportional trading charges.
Before the stamp-duty reduction that took effect on August 28, 2023,
the corresponding non-commission cost was approximately 5 basis points under the same convention \citep{MOFSTA2023StampDuty}.
Accordingly,
the 5- and 10-basis-point scenarios provide relevant benchmarks for explicit transaction charges over the 2021--2025 out-of-sample period,
while the 15-basis-point case constitutes a more stringent sensitivity test.

Table~\ref{tab:transaction_cost} reports annualized net returns and Sharpe ratios for one-way costs of 5, 10, and 15 basis points,
together with the gross results at zero cost.
Net Sharpe ratios are calculated from the cost-adjusted holding-period return series.
The significance of mean net returns is assessed using two-sided tests with Newey--West standard errors and one lag,
consistent with the procedure in Section~\ref{sec:portfolio_2DCNN}.

For the H--L portfolio, the annualized return declines from 0.38 before costs to 0.28 at 5 basis points, while the Sharpe ratio declines from 3.37 to 2.48.
At 10 basis points, the portfolio retains an annualized net return of 0.18 and a Sharpe ratio of 1.59, and the net return remains significant at the 1\% level.
At 15 basis points, however, the annualized net return falls to 0.08 and the Sharpe ratio to 0.69, and the net return is no longer statistically significant.
The break-even one-way cost is 18.86 basis points.

The High-only portfolio is less sensitive to transaction costs because it does not incur costs on a short leg.
Its annualized return and Sharpe ratio decline from 0.31 and 1.40 before costs to 0.26 and 1.17 at 5 basis points and to 0.21 and 0.95 at 10 basis points.
The net return remains significant at the 5\% level under the 10-basis-point assumption.
At 15 basis points,
the portfolio continues to earn a positive annualized net return of 0.16, with a Sharpe ratio of 0.73,
although the net return is no longer statistically significant.
Its break-even one-way cost is 31.43 basis points, substantially higher than that of the H--L portfolio.
\begin{table}[H]
	\centering
	\caption{\textbf{Transaction-cost-adjusted portfolio performance.}
	This table reports the out-of-sample performance of the equal-weighted High-only and H--L portfolios formed on forecasts from the OHLC-only 2D CNN
	after accounting for transaction costs.
	Ret and SR denote the annualized mean net return and annualized net-of-cost Sharpe ratio, respectively.
	Break-even cost denotes the one-way transaction cost at which the annualized net return becomes zero.
	***, **, and * denote significance at the 1\%, 5\%, and 10\% levels, respectively.}
	\label{tab:transaction_cost}
	\begin{tabular*}{\textwidth}{@{\extracolsep{\fill}} l *{2}{cc}}
		\toprule
		\multirow{2}{*}{Transaction cost}     & \multicolumn{2}{c}{High-only portfolio}    & \multicolumn{2}{c}{H--L portfolio} \\
		                                      \cmidrule(lr){2-3}                           \cmidrule(lr){4-5}
		                                      & Ret     & SR                               & Ret     & SR   \\
		\midrule
		0 basis points                        & 0.31*** & 1.40                             & 0.38*** & 3.37 \\
		5 basis points                        & 0.26*** & 1.17                             & 0.28*** & 2.48 \\
		10 basis points                       & 0.21**  & 0.95                             & 0.18*** & 1.59 \\
		15 basis points                       & 0.16    & 0.73                             & 0.08    & 0.69 \\
		Break-even cost                       & \multicolumn{2}{c}{31.43 basis points}     & \multicolumn{2}{c}{18.86 basis points} \\
		\bottomrule
	\end{tabular*}
\end{table}

Overall, proportional transaction costs materially reduce the profitability of both short-horizon strategies, with a particularly pronounced effect on the H--L portfolio.
Both portfolios retain positive and statistically significant net returns at one-way costs of up to 10 basis points,
although the magnitude and statistical significance of net performance weaken materially as costs rise.
Because the H--L cost adjustment does not incorporate securities-lending fees or constraints on borrow availability,
its net performance should be interpreted as a transaction-cost sensitivity benchmark rather than as the return of a fully implementable long--short strategy.
The greater cost tolerance of the High-only portfolio makes it a more relevant implementation benchmark in the Chinese market,
where establishing and maintaining the short leg is substantially more constrained.

\subsection{Trading constraints in the Chinese A-share market}
Beyond proportional transaction costs, implementation in the Chinese A-share market is affected by price limits and trading suspensions.
Because the short leg of the H--L portfolio is difficult to implement in practice, we focus on the High-only portfolio as the more relevant implementation benchmark.

Several constraints observable at portfolio formation are already addressed by the sample construction in Section~\ref{sec:data}.
All sample firms were listed before January 1, 2011, while the empirical period begins in 2013, thereby excluding trading conditions specific to newly listed stocks.
Candidate stock windows are excluded if any observation in the 20-day input window carries an ST designation; consequently,
stocks with ST observations in their input windows are not eligible for portfolio formation.
Suspensions occurring after formation, by contrast, are not known ex ante.

We first examine potential entry constraints caused by upper price limits.
Because our adjusted-price data do not contain official daily price-limit or order-execution indicators, we apply a conservative near-limit-up screen.
At each formation date, we first select the $N$ stocks with the highest OHLC-only 2D CNN probabilities,
remove those whose adjusted return on that date exceeds 9\%, and reweight the remaining stocks equally.
Table~\ref{tab:price_limit} reports results for $N=170$, 100, 50, and 10.
All four screened portfolios retain positive and statistically significant gross returns.
After imposing a one-way transaction cost of 10 basis points,
their annualized net returns remain between 0.18 and 0.21, with three of the four estimates statistically significant.
Thus, even after excluding stocks with formation-day adjusted returns above 9\%,
the screened long-only portfolios continue to generate economically meaningful returns and Sharpe ratios.
\begin{table}[H]
	\centering
	\caption{\textbf{Performance of Top-$N$ long-only portfolios after a near-limit-up screen.}
	This table reports the out-of-sample gross and net performance of Top-$N$ portfolios
	formed from the forecasts of the OHLC-only 2D CNN after excluding stocks whose adjusted return on the formation date exceeds 9\%.
	Net performance assumes a one-way transaction cost of 10 basis points.
	Ret and SR denote the annualized mean return and annualized Sharpe ratio, respectively,
	while Turnover denotes average monthly turnover.
	***, **, and * denote significance of mean returns at the 1\%, 5\%, and 10\% levels, respectively.}
	\label{tab:price_limit}
	\begin{tabular*}{\textwidth}{@{\extracolsep{\fill}} l*{5}{c}}
		\toprule
		\multirow{2}{*}{Top-$N$}  & \multirow{2}{*}{Turnover}    & \multicolumn{2}{c}{Gross performance}    & \multicolumn{2}{c}{Net performance} \\
		                                                         \cmidrule(lr){3-4}                         \cmidrule(lr){5-6}
		                          &                              & Ret     & SR                             & Ret     & SR   \\
		\midrule
		170                       & 333\%                        & 0.28*** & 1.28                           & 0.18*   & 0.83 \\
		100                       & 354\%                        & 0.29*** & 1.29                           & 0.19*   & 0.85 \\
		50                        & 369\%                        & 0.31*** & 1.33                           & 0.21**  & 0.90 \\
		10                        & 390\%                        & 0.30**  & 1.03                           & 0.20    & 0.69 \\
		\bottomrule
	\end{tabular*}
\end{table}

We next examine suspensions arising during the holding period.
Across the 239 out-of-sample holding periods, 48 contain at least one suspended High-portfolio stock, with 66 stock--period suspension instances in total.
Of these 48 affected periods, 42 contain only one suspended stock, and the maximum number in any single period is seven.
For an affected stock, suspended days do not advance its holding-day count;
the position is carried forward until trading resumes and the remaining active trading days are observed.
This treatment does not assume execution while the stock is suspended.
Nevertheless, because new portfolios continue to be formed on the common five-market-day schedule,
the backtest does not fully capture the opportunity cost of capital being locked in suspended positions or
the additional capital required to maintain these positions alongside newly formed portfolios.

For the same portfolio, we also assess potential exit constraints associated with lower price limits.
An adjusted daily return below $-9\%$ on the fifth subsequent active trading observation
is used as a proxy because the available data cannot identify whether a sell order was actually unfilled.
No such stock appears in 156 of the 239 periods, and no more than five appear in 234 periods, or 97.9\% of the sample.
Counts above five occur in only five periods, which account for 68.5\% of all such observations.
These concentrated episodes represent a salient form of tail risk,
as potential exit constraints that are limited under normal market conditions can become widespread during episodes of market-wide stress.

Taken together, the High-only results survive a conservative entry screen, while suspensions and potential lower-limit exposure usually affect only a small number of holdings.
The backtest nevertheless cannot fully reproduce delayed execution, capital lock-up, or unfilled orders during extreme market conditions.
In addition, market impact and liquidity constraints are not explicitly modeled, and strategy capacity is not assessed.
The long-only results should therefore be interpreted as evidence on the economic value of the predictive signal under a small-scale,
price-taking implementation rather than as a complete simulation of trading under all market frictions.

\section{Conclusion}
\label{sec:conclusion}
This study examines the standalone economic value of trading volume and three groups of manually constructed technical indicators and
whether adding these groups to OHLC candlestick charts provides incremental out-of-sample economic value.
The seven indicators are organized as follows:
the trend group comprises SMA, ADX together with its positive and negative directional components, and MACD;
the momentum group comprises RSI and STOCH;
and the volatility group comprises Bollinger Bands and ATR-based Bands.
Using daily data for 1,707 stocks listed on the main boards of the Chinese A-share market from 2013 to 2025,
we construct approximately 3.36 million stock-window samples from 20-day market histories,
encode each sample as both a multi-channel image and a numerical representation,
and use these representations to predict the direction of subsequent five-trading-day returns.
The primary analysis applies a 2D CNN to the multi-channel images,
while complementary tests apply an $\ell_2$-regularized logistic regression, a 1D CNN, and an LSTM to numerical representations of the same samples.
Economic performance over the 2021--2025 out-of-sample period is evaluated through equal-weighted decile portfolios,
H--L returns, Sharpe ratios, Fama--French factor-adjusted alphas, and formal pairwise tests.

The standalone results show that, under all four model architectures,
specifications based separately on OHLC, trading volume, and each technical-indicator group generate significantly positive H--L returns and factor-adjusted alphas,
together with economically meaningful Sharpe ratios.
The strength of the standalone advantage of OHLC nevertheless varies across architectures.
Under the 2D CNN and 1D CNN, the OHLC-only specifications significantly outperform the corresponding specifications based on trading volume or any technical-indicator group
in H--L returns, Sharpe ratios, and factor-adjusted alphas.
The OHLC-only LSTM also produces the largest point estimates, although its advantages are statistically significant in only some comparisons.
Under the $\ell_2$-regularized logistic regression, the corresponding performance differences are not statistically significant.
These results show that all standalone specifications generate economically meaningful portfolio performance,
while the OHLC-only specifications generally perform best under the nonlinear models considered here.

The evidence concerning incremental value is more consistent across architectures.
None of the augmented specifications produces a statistically significant increase in
either the mean H--L return or factor-adjusted alpha relative to OHLC alone under any of the four models.
The technical-indicator additions also provide no robust improvement in Sharpe ratios,
and some instead significantly reduce H--L returns, Sharpe ratios, and factor-adjusted alphas.
The only statistically significant improvements over OHLC arise in the LSTM Sharpe ratios:
OHLC + Volume and All features generate significantly higher Sharpe ratios than OHLC alone.
However, the All features specification does not significantly outperform OHLC + Volume in either return or Sharpe ratio.
This pattern is consistent with the LSTM Sharpe-ratio improvement being associated primarily with trading volume rather than with the technical-indicator groups.
For the 2D CNN, the cumulative volatility-adjusted log-return paths similarly provide no evidence that an augmented-input specification consistently dominates OHLC alone.
These results support a conclusion of limited incremental economic value for these indicator groups,
but do not establish that technical indicators are generally redundant.

The benchmark and implementation analyses using the baseline OHLC-only 2D CNN qualify the economic interpretation of these results.
The model produces a more orderly cross-sectional return ranking and significantly higher Sharpe ratios than the four traditional trading strategies,
although the statistical significance of its mean-return and alpha advantages varies across benchmarks.
At the same time, the high turnover of the model-based portfolios makes trading costs economically important.
Both the H--L and High-only portfolios retain positive and statistically significant net returns at proportional one-way costs of up to 10 basis points,
but their performance weakens materially as costs increase.
The High-only results also remain positive after a conservative near-limit-up screen.
Trading suspensions and potential lower-limit exposure usually affect only a small number of holdings,
although they can become concentrated during periods of market-wide stress.
The reported H--L net returns exclude securities-lending fees and do not account for constraints on borrow availability.
Because the backtest does not fully incorporate market impact, liquidity constraints,
delayed execution, or the opportunity cost of capital locked in suspended positions,
the long-only results should be interpreted as evidence on the economic value of the predictive signal under
a small-scale, price-taking implementation rather than as a complete simulation of trading under all market frictions.

Together, these results provide the first large-scale evidence from the Chinese A-share market on
the standalone and incremental economic value of the technical-indicator groups examined here across multiple representations and model architectures.
The broadly consistent evidence indicates that their limited incremental value is not specific to
the candlestick-image representation or the 2D CNN architecture.

The findings are specific to the trend, momentum, and volatility indicators examined here,
the four model architectures, and the Chinese main-board A-share sample.
They are also conditional on the indicator parameters,
scaling and estimation procedures,
and 20-day window and five-day prediction horizon used in this study.
Future research could assess whether the results extend to other technical-indicator specifications, market environments, and model architectures.
It could also examine additional inputs beyond OHLC, such as order-book variables, textual news, or measures of investor attention,
and investigate how these inputs interact with price-based representations.
Further research is also needed to clarify the mechanisms underlying model forecasts.
Feature-attribution methods and analyses of learned representations could be used to identify the temporal and spatial patterns that
drive portfolio rankings and to examine whether these patterns vary across market states.
Such analyses may help determine whether the models rely on trend persistence, reversal, volatility, or other price configurations.
Finally, more detailed transaction-level data could provide a fuller assessment of execution costs, strategy capacity, and performance under binding market constraints.

\section*{Declaration of competing interest}
The authors declare that they have no known competing financial interests or personal relationships that could have appeared to influence the work reported in this paper.

\section*{Data availability}
The data are available from the Baostock (\url{https://baostock.com}) and CSMAR (\url{https://data.csmar.com/}) databases,
or from the corresponding author upon reasonable request.

\section*{Acknowledgements}
This work was supported by
the Shaanxi Academy of Educational Sciences of China (grant number SGH25Y2666),
the National Natural Science Foundation of China (grant number 72501215),
and the Humanities and Social Sciences Youth Foundation of the Ministry of Education of China (grant number 23XJC630014).

\small{
	\bibliography{refs}
}

\end{document}